\documentclass{aa}
\usepackage{graphicx}
\usepackage{amssymb}
\usepackage{natbib}
\usepackage{enumerate}
\usepackage{enumitem}
\usepackage{float}
\usepackage{color}
\usepackage{longtable}
\usepackage{xfrac}
\usepackage[breaklinks]{hyperref}
\usepackage[varg]{txfonts}
\newcommand{\refeq}[1]{Eq.~(\ref{#1})}
\newcommand{\refeqs}[2]{Eqs.~(\ref{#1})--(\ref{#2})}
\newcommand{\refeqsand}[2]{Eqs.~(\ref{#1}) and~(\ref{#2})}
\newcommand{\reffig}[1]{Fig.~\ref{#1}}
\newcommand{\reftab}[1]{Table~\ref{#1}}
\newcommand{\refapp}[1]{Appendix~\ref{#1}}
\newcommand{\refsec}[1]{Sect.~\ref{#1}}
\newcommand{\ie}{i.e.\ }
\newcommand{\eg}{e.g.\ }
\newcommand{\kep}{{\it Kepler}}

\hypersetup{
    colorlinks=true,
    linkcolor=blue,
	citecolor=blue,
    filecolor=blue,      
	urlcolor=blue,
}
\begin{document}
%\modulolinenumbers[5]
%\linenumbers
%
   %;;;%\title{Asteroseismic modelling and orbital analysis of the triple stars HD~188753 %;;;%observed by \kep{}
   \title{Asteroseismic and orbital analysis of the triple star system HD~188753 observed by \kep{}
   }

%;;;%\subtitle{Implication for the dynamical evolution of the system}

   \author{F.~Marcadon
          %\inst{1}
          \and
          T.~Appourchaux
          %\inst{1}
          \and
          J.~P.~Marques
          %\inst{1}
          }

   \institute{Institut d'Astrophysique Spatiale, CNRS, Univ. Paris-Sud, Universit\'{e} Paris-Saclay, B\^{a}t. 121, 91405 Orsay cedex, France
          }

   \date{Version 1.0 -- 24 April 2018.}

% \abstract{}{}{}{}{} 
% 5 {} token are mandatory
 
  \abstract
  {The NASA \kep{} space telescope has detected solar-like oscillations in several hundreds of single stars, thereby providing a way to determine precise stellar parameters using asteroseismology.}
   {In this work, we aim to derive the fundamental parameters of a close triple star system, HD~188753, for which asteroseismic and astrometric observations allow independent measurements of stellar masses.}
   {We used six months of \kep{} photometry available for HD~188753 to detect the oscillation envelopes of the two brightest stars. For each star, we extracted the individual mode frequencies by fitting the power spectrum using a maximum likelihood estimation approach. We then derived initial guesses of the stellar masses and ages based on two seismic parameters and on a characteristic frequency ratio, and modelled the two components independently with the stellar evolution code CESTAM. In addition, we derived the masses of the three stars by applying a Bayesian analysis to the position and radial-velocity measurements of the system.}
   {Based on stellar modelling, the mean common age of the system is $10.8 \pm 0.2\,{\rm Gyr}$ and the masses of the two seismic components are $M_A =$ $0.99 \pm 0.01\,M_\odot$ and $M_{Ba} =$ $0.86 \pm 0.01\,M_\odot$. From the mass ratio of the close pair, $M_{Bb}/M_{Ba}  = 0.767 \pm 0.006$, the mass of the faintest star is $M_{Bb} =$ $0.66 \pm 0.01\,M_\odot$ and the total seismic mass of the system is then $M_{\rm syst} =$ $2.51 \pm 0.02\,M_\odot$. This value agrees perfectly with the total mass derived from our orbital analysis, $M_{\rm syst} =$ $2.51^{+0.20}_{-0.18}\,M_\odot$, and leads to the best current estimate of the parallax for the system, $\pi = 21.9 \pm 0.2\,{\rm mas}$. In addition, the minimal relative inclination between the inner and outer orbits is $10.9^\circ \pm 1.5^\circ$, implying that the system does not have a coplanar configuration.}
   {}

   \maketitle

%%%%%%%%%%%%%%%%%%%%%%%%%%%%%%%%%%%%%%%%%%%%%%%%%%%%%%%%%%%%%
%%%%%%%%%%%%%%%%%%%%%%%%%%Introduction%%%%%%%%%%%%%%%%%%%%%%%%%%%
%%%%%%%%%%%%%%%%%%%%%%%%%%%%%%%%%%%%%%%%%%%%%%%%%%%%%%%%%%%%%

\section{Introduction}

Stellar physics has experienced a revolution in recent years with the success of the  CoRoT \citep{2006ESASP1306...33B} and \kep{} space missions \citep{2010PASP..122..131G}. Thanks to the high-precision photometric data collected by CoRoT and \kep{}, asteroseismology has matured into a powerful tool for the characterisation of stars.

The \kep{} space telescope yielded unprecedented data allowing the detection of solar-like oscillations in more than 500 stars \citep{2011Sci...332..213C} and the extraction of mode frequencies for a large number of targets \citep{2012A&A...543A..54A,2016MNRAS.456.2183D,2017ApJ...835..172L}. From the available sets of mode frequencies, several authors performed detailed modelling to infer the mass, radius, and age of stars \citep{2014ApJS..214...27M,2015MNRAS.452.2127S,2017A&A...601A..67C}. In addition, using scaling relations, the measurement of the seismic parameters $\Delta\nu$ and $\nu_{\rm max}$ provides a model-free estimate of stellar mass and radius \citep{2011Sci...332..213C,2014ApJS..210....1C}. A direct measurement of stellar mass and radius can also be obtained using the large frequency separation, $\Delta\nu$, angular diameter from interferometric observations, and parallax (see \eg \citealt{2012ApJ...760...32H,2013MNRAS.433.1262W}). The determination of accurate stellar parameters is crucial for studying the populations of stars in our Galaxy \citep{2011Sci...332..213C,2013MNRAS.429..423M}. Therefore, a proper calibration of the evolutionary models and scaling relations is required in order to derive the stellar mass, radius, and age with a high level of accuracy. In this context, binary stars provide a unique opportunity to check the consistency of the derived stellar quantities.

Most stars are members of binary or multiple stellar systems. Among all the targets observed by \kep{}, there should be many systems showing solar-like oscillations in both components, that is, seismic binaries. However, using population synthesis models, \cite{2014ApJ...784L...3M} predicted that only a small number of seismic binaries are expected to be detectable in the \kep{} database. Indeed, the detection of the fainter star requires a magnitude difference between both components typically smaller than approximately one. Until now, only four systems identified as seismic binaries using \kep{} photometry have been reported in the literature: 16~Cyg~A and B (KIC~12069424 and KIC~12069449; \citealt{2012ApJ...748L..10M,2015ApJ...811L..37M,2015MNRAS.446.2959D}), HD~176071 (KIC~9139151 and KIC~9139163; \citealt{2012A&A...543A..54A,2014ApJS..214...27M}), HD~177412 (KIC~7510397; \citealt{2015A&A...582A..25A}) and HD~176465 (KIC~10124866; \citealt{2017A&A...601A..82W}).

The theory of binary star formation excluded the gravitational capture mechanism as an explanation for the existence of binary systems \citep{Tohline1992}. As a consequence,  the binary star systems are created from the same dust disk during stellar formation \citep{Tohline1992}. Due to their common origin, both stars of a seismic binary are assumed to have the same age and initial chemical composition, allowing a proper calibration of stellar models. Indeed, the detailed modelling of each star results in two independent values of the stellar age, which can be compared a posteriori. In addition, ground-based observations of a seismic binary also provide strong constraints on the stellar masses through the dynamics of the system. For example, the total mass of the system can be derived from its relative orbit using interferometric measurements. Among the four seismic binaries quoted above, only the pair HD~177412 has a sufficiently short period, namely $\sim$14$\,{\rm yr}$, to allow the determination of an orbital solution for the system. From the semi-major axis and the period of the relative orbit, \cite{2015A&A...582A..25A} derived the total mass of the system using Kepler's third law associated with the revised Hipparcos parallax of \cite{2007AA...474..653V}. However, the parallax of such a binary star can be affected by the orbital motion of the system \citep{2008IAUS..248...59P}, resulting in a potential bias on the estimated values of total mass and stellar radii.

Another approach for measuring stellar masses is to combine spectroscopic and interferometric observations of binary stars. This method has the advantage of allowing a direct determination of distance and individual masses. Such an analysis has recently been performed by \cite{2016A&A...586A..90P} for the well-known binary $\alpha$~Cen~AB, which shows solar-like oscillations in both components (\citealt{2002A&A...390..205B,2003A&A...406L..23C}). Thus far, $\alpha$~Cen~AB is the only seismic binary for which both a direct comparison between the estimated values of the stellar masses, using asteroseismology and astrometry, and an independent distance measurement are possible. Fortunately, another system offers the possibility of performing such an analysis using \kep{} photometry and ground-based observations. This corresponds to a close triple star system, \object{HD 188753}, for which we have detected solar-like oscillations in the two brightest components.

This article is organised as follows. \refsec{sec:obs} summarises the main features of HD~188753 and briefly describes the observational data used in this work, including \kep{} photometry and ground-based measurements. \refsec{sec:orb_analysis} presents the orbital analysis of HD~188753 leading to the determination of the distance and individual masses. \refsec{sec:seismic} details the seismic analysis of the two oscillating components, providing accurate mode frequencies and reliable proxies of the stellar masses and ages. \refsec{sec:modelling} describes the input physics and optimisation procedure used for the detailed modelling of each of the two stars. Finally, the main results of this paper are presented and discussed in \refsec{sec:results}, and some conclusions are drawn in \refsec{sec:conclusions}.

%%%%%%%%%%%%%%%%%%%%%%%%%%%%%%%%%%%%%%%%%%%%%%%%%%%%%%%%%%%%%
%%%%%%%%%%%%%%%%%%%%%%%%%%Target and observations%%%%%%%%%%%%%%%%%%%%%%%%%%%
%%%%%%%%%%%%%%%%%%%%%%%%%%%%%%%%%%%%%%%%%%%%%%%%%%%%%%%%%%%%%

\section{Target and observations}\label{sec:obs}

%%%%%%%%%%%%%%%%%%%%%%%%%%Time series and power spectrum%%%%%%%%%%%%%%%%%%%%%%%%%%%

\subsection{Time series and power spectrum}

HD~188753 is a bright triple star system ($V =$ $7.41\,{\rm mag}$) situated at a distance of $45.7 \pm 0.5\,{\rm pc}$. The main stellar parameters of the system are given in \reftab{tab:stellar_param}. This system was observed by the \kep{} space telescope in short-cadence mode (58.85 s sampling; \citealt{2010ApJ...713L.160G}) during a time period of six months, between 2012 March~29 and 2012 October~3. Standard \kep{} apertures were not designed for the observation of such a bright and saturated target. A custom aperture was therefore defined and used for HD~188753, through the \kep{} Guest Observer (GO) program\footnote{\url{https://keplergo.arc.nasa.gov/}}, to capture all of the stellar flux.

\begin{table}[tbp]
\caption{Main stellar and seismic parameters of HD~188753.}
\centering
\begin{tabular}{l c c} 
\hline
\hline
Stellar parameter &  HD~188753 & Reference \\
\hline
$V$ (mag) & 7.41 & \cite{1997ESASP1200.....E}\\
$\Delta V$\tablefootmark{a} (mag) & 0.69 & \cite{1997ESASP1200.....E} \\
$T_{\rm eff}$ (K) & $5383 \pm 188$ & \cite{2014ApJS..211....2H} \\
$\log g$ (cgs) & $4.52 \pm 0.40$ & \cite{2014ApJS..211....2H} \\
\rm{[Fe/H]} (dex) & $0.06 \pm 0.30$ & \cite{2014ApJS..211....2H} \\
$V_0$ (km s$^{-1}$) & $-22.21 \pm 0.03$ & our orbital estimate\tablefootmark{b} \\
Parallax (mas) & $21.9 \pm 0.6$ & our orbital estimate\tablefootmark{b} \\
Parallax (mas) & $21.9 \pm 0.2$ & our seismic estimate\tablefootmark{c} \\
Distance (pc) & $45.7 \pm 0.5$ & our seismic estimate\tablefootmark{c} \\
\hline
Seismic parameter & Star~A & Star~Ba \\
\hline
$\nu_{\rm max}$ ($\mu$Hz) & $2204 \pm 8$ & $3274 \pm 67$ \\
$\Delta\nu$ ($\mu$Hz) & $106.96 \pm 0.08$ & $147.42 \pm 0.08$ \\
\hline
\end{tabular}
\tablefoot{%
\tablefoottext{a}{Magnitude difference between the visual components.}
\tablefoottext{b}{Systemic velocity and parallax derived in \refsec{sec:orbAB}.}
\tablefoottext{c}{Parallax and corresponding distance derived in \refsec{sec:parallax}.}
}
\label{tab:stellar_param}
\end{table}

Short-cadence (SC) photometric data are available to the \kep{} Asteroseismic Science Consortium (KASC; \citealt{2010AN....331..966K}) through the \kep{} Asteroseismic Science Operations Center (KASOC) database\footnote{\url{http://kasoc.phys.au.dk/}}. The SC time series for HD~188753 (KIC~6469154) is divided into quarters of three months each, referred to as Q13 and Q14. Custom Aperture File (CAF) observations are handled differently from standard \kep{} observations. A specific \kep{} catalogue ID\footnote{\kep{} IDs are $100\,004\,033$ and $100\,004\,071$ for Q13 and Q14 respectively.} greater than $100\,000\,000$ is then assigned to each quarter. This work is based on SC data in Data Release 25 (DR25)\footnote{DR25 corresponds to version 4 of Q13-Q14 for HD~188753.}, which were processed with the SOC Pipeline 9.3 \citep{2010ApJ...713L..87J}. As a result, the DR25 SC light curves were corrected for a calibration error affecting the SC pixel data in DR24.

\begin{figure*}[tbp]
\centering
\vbox{
\includegraphics[trim = 0.5cm 0.cm 1.cm 1.cm, clip,width=0.35\textwidth ,angle=90]{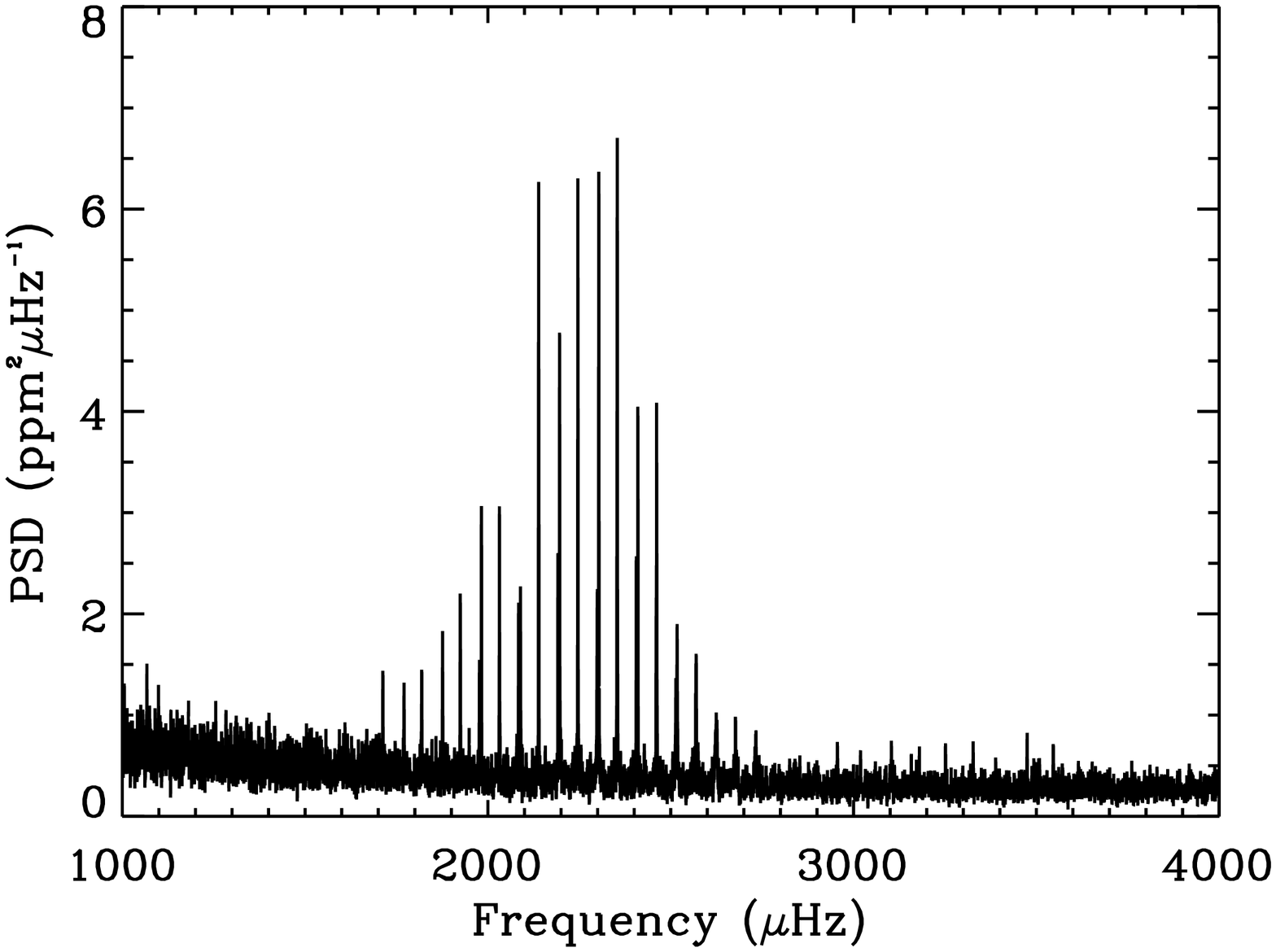}
\includegraphics[trim = 0.5cm 0.cm 1.cm 1.cm, clip,width=0.35\textwidth ,angle=90]{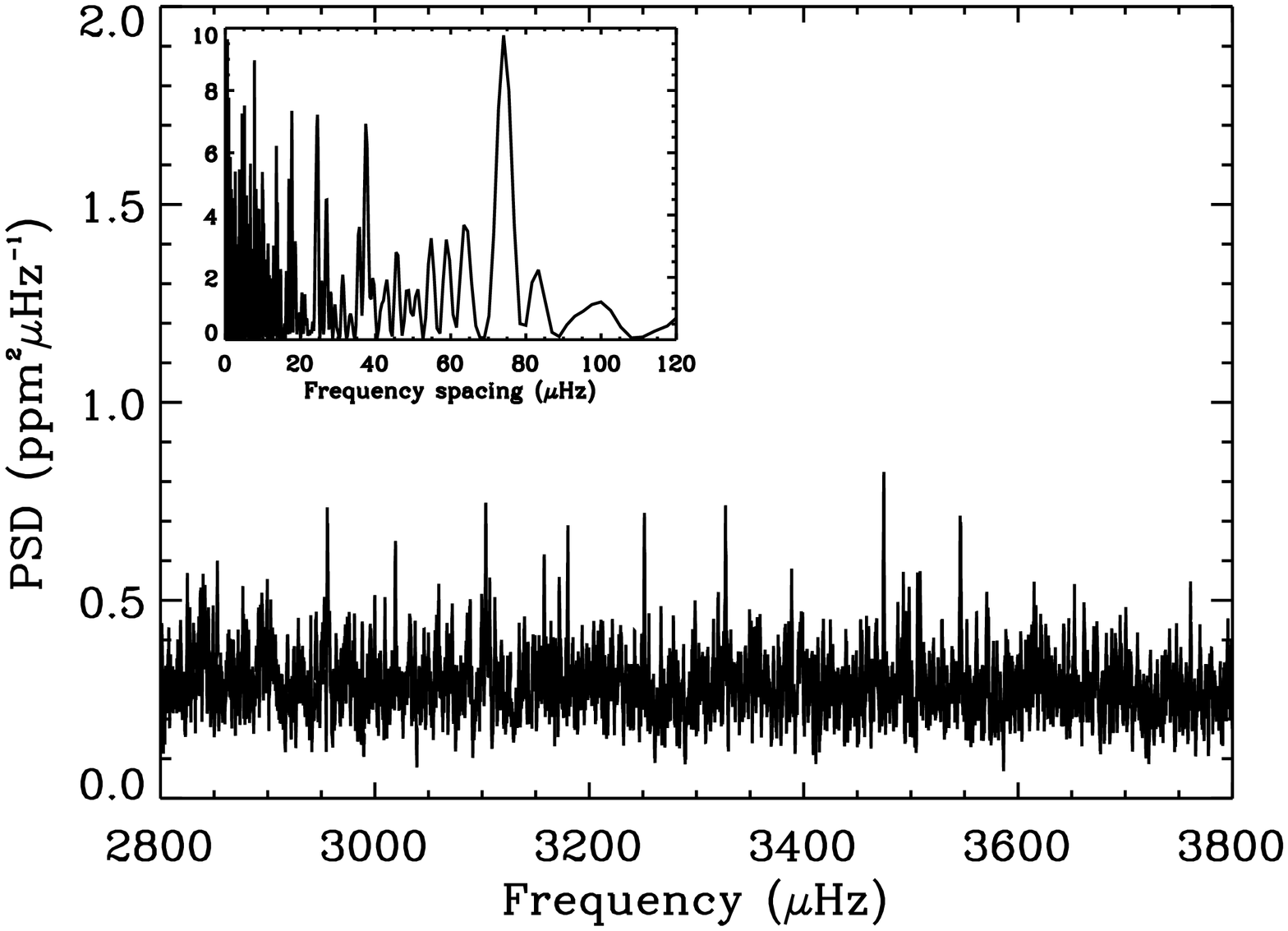}
% for a referee version
%\includegraphics[trim = 0.5cm 0.cm 1.cm 1.cm, clip,width=0.34\textwidth ,angle=90]{PSPS_6469154_sm15_compA.ps}
%\includegraphics[trim = 0.5cm 0.cm 1.cm 1.cm, clip,width=0.34\textwidth ,angle=90]{PSPS_6469154_sm15_compBa.ps}
}
\caption{Power spectrum of HD~188753. Left panel: Power spectrum smoothed with a $1\,\mu{\rm Hz}$ boxcar filter showing the oscillation modes of the two seismic components around $2200\,\mu{\rm Hz}$ and $3300\,\mu{\rm Hz}$, respectively. Right panel: Zoom-in on the mode power peaks of the secondary component. The inset displays the power spectrum of the power spectrum (PSPS) computed over the frequency range of the oscillations around $\nu_{\rm max} = 3274 \pm 67\,\mu{\rm Hz}$ (see \reftab{tab:stellar_param}).}
\label{fig:PSPS_6469154}
\end{figure*}

The light curves were concatenated and high-pass filtered using a triangular smoothing with a full width at half maximum (FWHM) of one day to minimise the effects of long-period instrumental drifts. The single-sided power spectrum was produced using the Lomb-Scargle periodogram \citep{1976Ap&SS..39..447L,1982ApJ...263..835S}, which has been properly calibrated to comply with Parseval’s theorem (see \citealt{2014aste.book..123A}). The length of data gives a frequency resolution of about $0.06\,{\rm \mu Hz}$. Figure~\ref{fig:PSPS_6469154} shows the smoothed power spectrum of HD~188753 with a zoom-in on the oscillation modes of the secondary seismic component at $\sim$3300$\,\mu{\rm Hz}$. The significant peak in the PSPS lies at $\Delta\nu /2\simeq 74\,\mu{\rm Hz}$, where $\Delta\nu$ is the large frequency separation; it corresponds to the signature of the near-regular spacing between individual modes of oscillation for the secondary component. \reftab{tab:stellar_param} provides the seismic parameters of the two stars, $\Delta\nu$ and $\nu_{\rm max}$, as derived in \refsec{sec:scal_rel}. We note that previous DR24 did not allow the detection of the secondary component in the SC light curves due to the degraded signal-to-noise ratio (S/N).

%%%%%%%%%%%%%%%%%%%%%%%%%%Astrometric and radial-velocity data%%%%%%%%%%%%%%%%%%%%%%%%%%%

\subsection{Astrometric and radial-velocity data}

HD~188753 (HIP~98001, HO~581 or WDS~19550+4152) is known as a close visual binary discovered by \cite{1899AN....149...65H} and characterised by an orbital period of $\sim$25 years \citep{1918AJ.....31..169V}. In the late 1970s, it was established that the secondary component is itself a spectroscopic binary with an orbital period of $\sim$154 days \citep{1977Obs....97...15G}. HD~188753 is thus a hierarchical triple star system consisting of a close pair~(B) in orbit at a distance of 11.8 AU from the primary component~(A).

\begin{figure}[tbp]
\centering
\includegraphics[trim = 0.5cm 5.5cm 1.cm 1.cm, clip,width=0.45\textwidth ,angle=90]{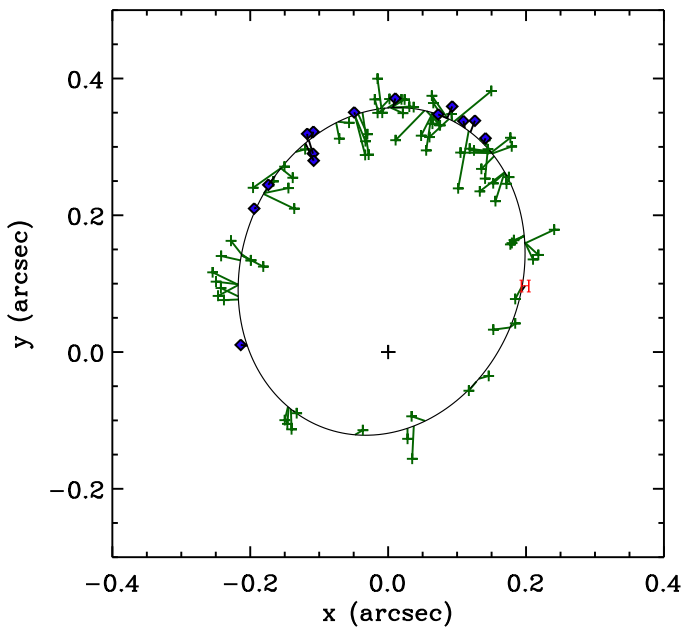}
% for a referee version
%\includegraphics[trim = 0.5cm 5.5cm 1.cm 1.cm, clip,width=0.70\textwidth ,angle=90]{VB_6469154_orbit.ps}
\caption{Astrometric orbit of HD~188753. Micrometric observations are indicated by green plus symbols and interferometric observations by blue diamonds.  The green lines indicate the distance between the observations and the fitted orbit. A red `H' indicates the Hipparcos measure. East is upwards and north is to the right.}
\label{fig:orbit_6469154}
\end{figure}

Position measurements of HD~188753 were first obtained by \cite{1899AN....149...65H} with the $18\,\sfrac{1}{2}$ inch Refractor of the Dearborn Observatory of Northwestern University. This system was then continuously observed over the last century using different techniques and instruments. 
%Visual observations (micrometry) are indicated by plus signs, interferometric observations (speckle or other single-aperture techniques) by filled diamonds and Hipparcos measures by the letter `H'. 
Since the pioneering work of \cite{1837sdmm.book.....S}, more than $150\,{\rm years}$ of micrometric measurements are now available for double stars in the literature. \reftab{tab:micro_data} provides the result of the micrometric observations for HD~188753. In addition, speckle interferometry for getting the relative position of close binaries has been in use since the 1970’s \citep{1974ApJ...194L.147L}. These observations are included in the Fourth Catalog of Interferometric Measurements of Binary Stars \citep{2001AJ....122.3480H}\footnote{\url{http://ad.usno.navy.mil/wds/int4.html}}. \reftab{tab:speckle_data} summarises the high-precision data obtained for HD~188753 from 1979 to 2006.  All published data of HD~188753 are displayed in \reffig{fig:orbit_6469154}.

The first radial-velocity observations of HD~188753 were performed by \cite{1977Obs....97...15G} from 1969 to 1975 using the photoelectric radial-velocity spectrometer of the Cambridge Observatories described in \cite{1967ApJ...148..465G}. The author then discovered that the system contained a spectroscopic binary with an orbital period of $\sim$154 days. HD~188753 was later observed by \cite{2005Natur.436..230K} with the high-resolution echelle spectrograph (HIRES; \citealt{1994SPIE.2198..362V}) at the W.~M.~Keck Observatory, from 2003 August to 2004 November. \citeauthor{2005Natur.436..230K} reported the detection of a hot Jupiter around the primary component that was challenged by \cite{2007A&A...466.1179E}. However, the author showed that the spectroscopic binary detected by \cite{1977Obs....97...15G} corresponds to the secondary component of the visual pair. The system was finally observed by \cite{2007A&A...466.1179E} with the ELODIE echelle spectrograph \citep{1996A&AS..119..373B} at the Observatoire de Haute-Provence (France) between July 2005 and August 2006. In order to derive the radial velocities of the faintest star, \cite{2009MNRAS.399..906M} applied a three-dimensional (3D) correlation technique (TRIMOR) to the data obtained by \cite{2007A&A...466.1179E}. In this study, we used the radial-velocity measurements from \cite{1977Obs....97...15G}, \cite{2005Natur.436..230K} and \cite{2009MNRAS.399..906M}.

%%%%%%%%%%%%%%%%%%%%%%%%%%%%%%%%%%%%%%%%%%%%%%%%%%%%%%%%%%%%%
%%%%%%%%%%%%%%%%%%%%%%%%%%Orbital analysis%%%%%%%%%%%%%%%%%%%%%%%%%%%
%%%%%%%%%%%%%%%%%%%%%%%%%%%%%%%%%%%%%%%%%%%%%%%%%%%%%%%%%%%%%

\section{Orbital analysis}\label{sec:orb_analysis}

%%%%%%%%%%%%%%%%%%%%%%%%%%Method and model%%%%%%%%%%%%%%%%%%%%%%%%%%%

\subsection{Method and model}\label{sec:orb_meth}

In this section, we present the methodology employed to determine the orbital parameters of the triple star system HD~188753. For this, we performed a combined treatment of more than a century of archival astrometry (AM) along with the radial-velocity (RV) measurements found in the literature.

Recently, \cite{2015A&A...582A..25A} derived the orbit of the seismic binary HD~177412 (HIP~93511) by applying a Bayesian analysis to the astrometric measurements of the system. We adapted this Bayesian approach in order to include the radial-velocity measurements for HD~188753. We defined the global likelihood of the data given the orbital parameters as:
\begin{equation}
{\cal L}={\cal L}_{\rm AM} \, {\cal L}_{\rm RV},\label{eq:lik_global}
\end{equation}
where ${\cal L}_{\rm AM}$ and ${\cal L}_{\rm RV}$ are the likelihoods of the AM and RV data respectively, computed from:
\begin{equation}
\ln {\cal L}_{\rm AM}= -\frac{1}{2}\,\sum^{N_{\rm AM}}_{i=1} \, \left(\left(\frac{x_i^{\rm mod}-x_i^{\rm obs}}{\sigma_{x,i}}\right)^2+\left(\frac{y_i^{\rm mod}-y_i^{\rm obs}}{\sigma_{y,i}}\right)^2\right),\label{eq:lik_AM}
\end{equation}
\begin{equation}
\ln {\cal L}_{\rm RV} = -\frac{1}{2}\,\sum^{N_{\rm RV}}_{i=1} \, \left(\frac{V_i^{\rm mod}-V_i^{\rm obs}}{\sigma_{V,i}} \right)^2.\label{eq:lik_RV}
\end{equation}
$N_{\rm AM}$ and $N_{\rm RV}$ denote the number of available AM and RV observations, respectively, and $\sigma$ refers to the associated uncertainties. The terms $x$ and $y$ correspond to the coordinates of the orbit on the plane of the sky and denote the declination and right ascension differences, respectively. The term $V$ stands for the radial velocities. The exponents `mod' and `obs' refer to the modelled and observed constraints used during the fitting procedure. Here, the observed positions $x^{\rm obs}$ and $y^{\rm obs}$ are computed from the measured quantities ($\rho$, $\theta$) by means of the simple relations $x=\rho\cos\theta$ and $y=\rho\sin\theta$, where $\rho$ is the relative separation and $\theta$ is the position angle for both components. \refapp{app:orb-data} provides the values of $\rho$ and $\theta$ derived from the micrometric and interferometric measurements of HD~188753. In addition, the radial-velocity observations $V^{\rm obs}$ used in this work can be found in \cite{1977Obs....97...15G}, \cite{2005Natur.436..230K} and \cite{2009MNRAS.399..906M}. The theoretical values in \refeqsand{eq:lik_AM}{eq:lik_RV}, namely $x^{\rm mod}$, $y^{\rm mod}$ and $V^{\rm mod}$, are calculated from the orbital parameters of the system following the observable model described in \refapp{app:obs-mod}. For the derivation of these orbital parameters, that is, ${\cal P}_{\rm orb}=(P,T,V_0,K_A,K_B,e,\omega,a,i,\Omega)$, we employed a Markov chain Monte Carlo (MCMC) method using the Metropolis-Hastings algorithm (MH; \citealt{1953JChPh..21.1087M,hastings70}) as explained in \refapp{app:bay-app}.

An essential aspect when combining different data types is the determination of a proper relative weighting. To this end, we performed a preliminary fit of each data set either coming from astrometry or from RV measurements, and calculated the root mean square (rms) of the residuals. \refapp{app:orb-param} provides the residual rms derived from the RV~measurements of the different authors. In the same way, we derived the residual rms on $x$ and $y$ by fitting the micrometric and interferometric measurements independently. For micrometric measurements, we found $\sigma_x =$ $25\,{\rm mas}$ and $\sigma_y =$ $34\,{\rm mas}$ while for interferometric measurements, we found $\sigma_x =$ $6\,{\rm mas}$ and $\sigma_y =$ $11\,{\rm mas}$. We used these values determined after the fitting as weights for the combined fit. 

%%%%%%%%%%%%%%%%%%%%%%%%%%Orbit of the AB system%%%%%%%%%%%%%%%%%%%%%%%%%%%

\subsection{Orbit of the AB system}\label{sec:orbAB}

We applied the methodology described above in order to derive the best-fit orbital solution of the visual pair, hereafter referred to as the AB system. However, the radial-velocity measurements need to be reduced from the 154-day modulation induced by the close pair, hereafter referred to as the Bab sub-system. We then fitted this 154-day modulation as described in \refsec{sec:orbit_Bab} (see also \refapp{app:orb-param}) to obtain the long-period orbital motion of the AB system.

\begin{figure*}[tbp]
\centering
\vbox{
\includegraphics[trim = -1.0cm 0.0cm -0.2cm -0.1cm, clip,width=0.38\textwidth ,angle=90]{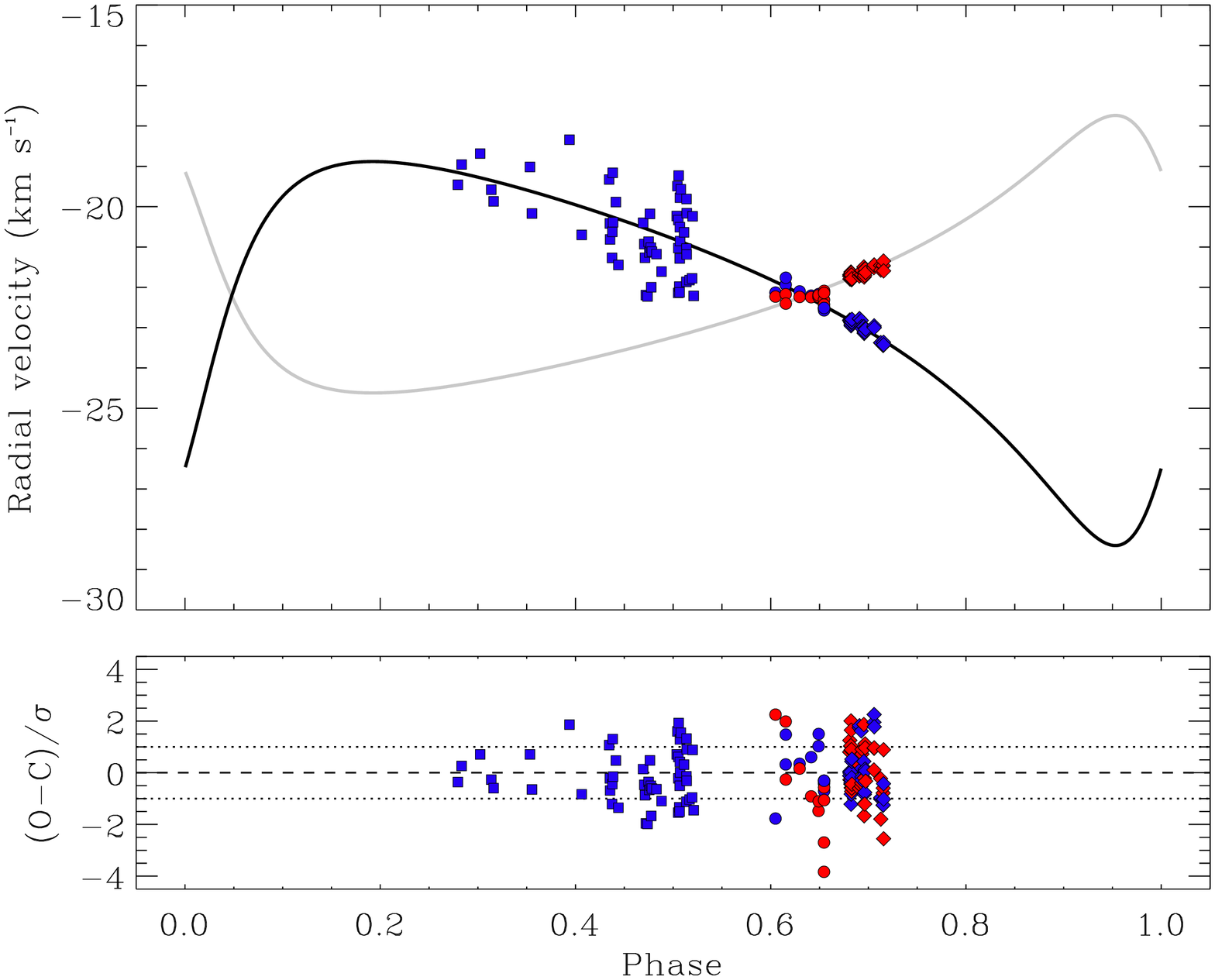}
\includegraphics[trim = -1.0cm 0.0cm -0.2cm -0.1cm, clip,width=0.38\textwidth ,angle=90]{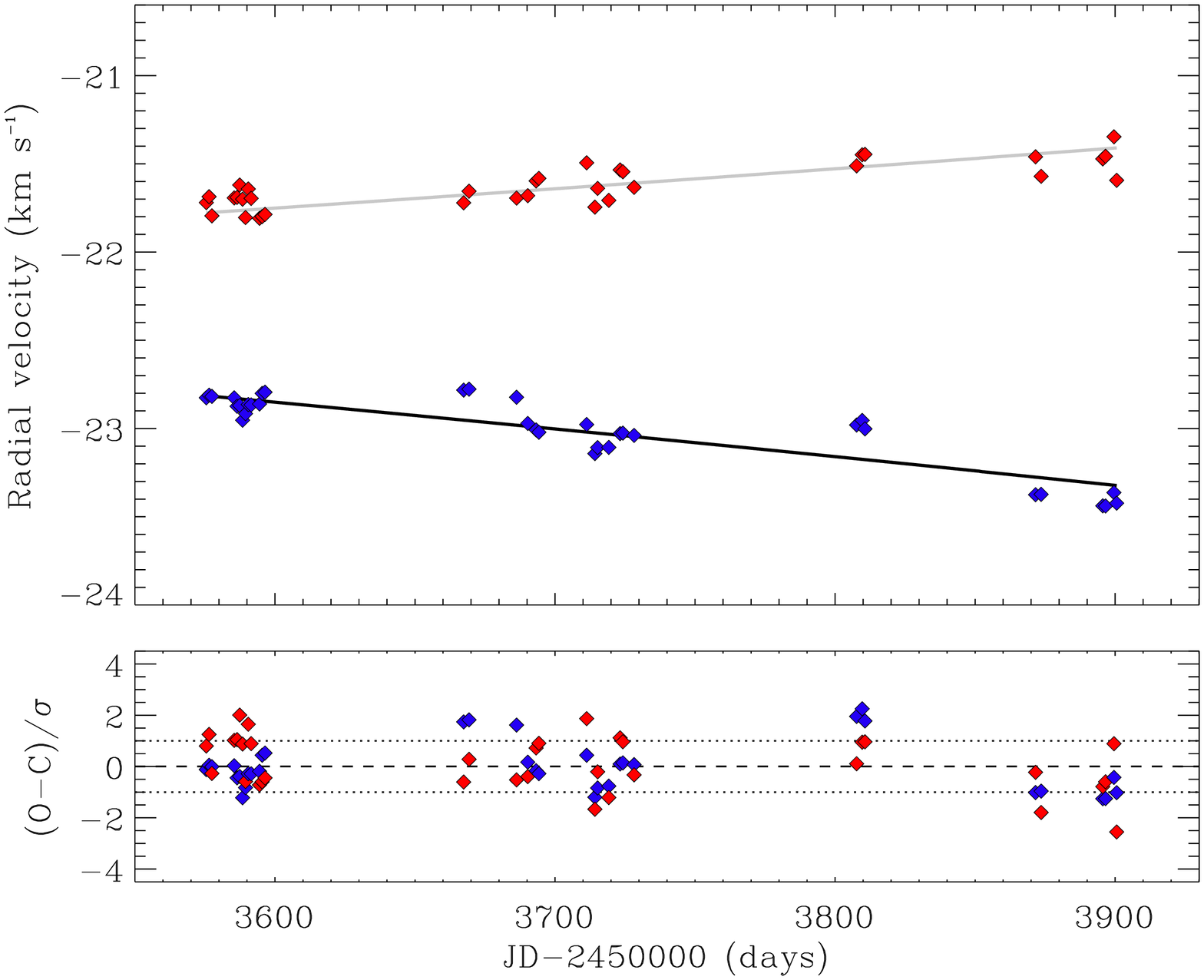}
}
\caption{Radial velocities of the two visual components. Left panel: RV measurements of the primary (blue) and secondary (red) components after having removed the 154-day modulation. Squares, circles, and diamonds denote the radial velocities from \cite{1977Obs....97...15G}, \cite{2005Natur.436..230K} and \cite{2009MNRAS.399..906M}, respectively. Right panel: Zoom-in on the RV measurements of \cite{2009MNRAS.399..906M} to show the quality of the fit. The black line denotes the RV solution of the A component while the grey line denotes the RV solution of the B component.}
\label{fig:RV_6469154}
\end{figure*}

Figure~\ref{fig:orbit_6469154} shows the best-fit solution of the astrometric orbit, plotted with all micrometric and interferometric data listed in \refapp{app:orb-data}.
The position of the primary component is marked by a plus symbol at the origin of the axes. $O-C$ residuals of the orbital solution are indicated by the solid lines connecting each observation to its predicted position along the orbit. We note that this orbit is listed as grade~1 in the Sixth Catalog of Orbits of Visual Binary Stars \citep{2001AJ....122.3472H}\footnote{\url{http://ad.usno.navy.mil/wds/orb6.html}}, on a scale of 1 (`definitive') to 5 (`indeterminate'), corresponding to a well-distributed coverage exceeding one revolution. In \reffig{fig:RV_6469154}, we also plotted the radial velocities of the two visual components computed from our best-fit solution. The black line denotes the RV~solution of the A~component, \ie the single star, while the grey line denotes the RV~solution of the Bab~sub-system. For comparison, we added the RV measurements from \cite{1977Obs....97...15G}, \cite{2005Natur.436..230K} and \cite{2009MNRAS.399..906M}, after having removed the 154-day modulation. The final orbital parameters determined from the best-fit solution are given in \reftab{tab:orbAB} .

\begin{table}[tbp]
\caption{Orbital and physical parameters of the AB system.}
\centering
\begin{tabular}{l c c c} 
\hline
\hline
Parameter & Median & 84$\%$ interval & 16$\%$ interval \\
\hline
$P$ (years) & 25.63 & +0.04 & -0.04 \\
$T$ (years) & 1988.1 & +0.1 & -0.1 \\
$V_0$ (km s$^{-1}$) & -22.21 & +0.03 & -0.03 \\
$K_A$ (km s$^{-1}$) & 4.8 & +0.2 & -0.2 \\
$K_B$ (km s$^{-1}$) & 3.4 & +0.3 & -0.3 \\
$e$ & 0.502 & +0.008 & -0.008 \\
$\omega$ (degrees) & 233.3 & +1.3 & -1.3 \\
$a$ (mas)\tablefootmark{a} & 258.7 & +2.8 & -2.7 \\
$i$ (degrees) & 31.1 & +1.5 & -1.5 \\
$\Omega$ (degrees) & 45.7 & +1.1 & -1.1 \\
\hline
$M_{A}$ ($M_\odot$) & 1.05 & +0.10 & -0.09 \\
$M_{B}$ ($M_\odot$) & 1.46 & +0.13 & -0.11 \\
$a$ (AU) & 11.8 & +0.3 & -0.3 \\
$\pi$ (mas)\tablefootmark{b} & 21.9 & +0.6 & -0.6 \\
\hline
\end{tabular}
\tablefoot{%
\tablefoottext{a}{mas is milliarcsecond.}
\tablefoottext{b}{Orbital parallax derived in \refsec{sec:orbAB}.}
}
\label{tab:orbAB}
\end{table}

From these parameters, we can derive the mass of the two components and the semi-major axis of the visual orbit \citep{1978GAM....15.....H}: 
%M_{A} =\frac{1}{(2\pi)^3}\,\frac{(K_A+K_B)^2\,K_B\,P\,(1-e^2)^{3/2}}{\sin^3 i}\label{eq:massA}
\begin{equation}
M_{A} = \frac{1.036 \times 10^{-7} (K_A+K_B)^2\,K_B\,P\,(1-e^2)^{3/2}}{\sin^3 i},\label{eq:massA}
\end{equation}
%M_{B} =\frac{1}{(2\pi)^3}\,\frac{(K_A+K_B)^2\,K_A\,P\,(1-e^2)^{3/2}}{\sin^3 i}\label{eq:massB}
\begin{equation}
M_{B} = \frac{1.036 \times 10^{-7} (K_A+K_B)^2\,K_A\,P\,(1-e^2)^{3/2}}{\sin^3 i},\label{eq:massB}
\end{equation}
%a _{\rm AU} = \frac{1}{2\pi}\,\frac{(K_A+K_B)\,P\,(1-e^2)^{1/2}}{\sin i}\label{eq:axisAB}
\begin{equation}
a _{\rm AU} = \frac{9.192 \times 10^{-5} (K_A+K_B)\,P\,(1-e^2)^{1/2}}{\sin i},\label{eq:axisAB}
\end{equation}
where $M_A$ and $M_B$ are expressed in units of the solar mass and $a _{\rm AU}$ is expressed in astronomical units. Here, $K_A$ and $K_B$ are the semi-amplitudes of the radial velocities for both components, in~km~s$^{-1}$, $P$~is the orbital period, in days, $e$~is the eccentricity and $i$~is the inclination of the plane of the orbit to the plane of the sky. In addition, the parallax of the system can be determined from the following ratio by combining the AM and RV results:
\begin{equation}
\pi_{\rm mas} = \frac{a_{\rm mas}}{a _{\rm AU}},\label{eq:parallax}
\end{equation}
where $\pi_{\rm mas}$ is expressed in milliarcseconds. Here, $a_{\rm mas}$ denotes the angular semi-major axis derived from the fitting procedure and given in \reftab{tab:orbAB} while $a _{\rm AU}$ denotes the linear semi-major axis defined in \refeq{eq:axisAB}. \reftab{tab:orbAB} provides the results of the above equations. In order to derive the median and the credible intervals of each physical parameter, namely $M_{A}$, $M_{B}$, $a _{\rm AU}$ and $\pi_{\rm mas}$, we calculated \refeqs{eq:massA}{eq:parallax} using the chains of the orbital parameters.

We found the orbital masses for the single star and the Bab sub-system to be $M_A =$ $1.05^{+0.10}_{-0.09}\,M_\odot$ and $M_B =$ $1.46^{+0.13}_{-0.11}\,M_\odot$, respectively. The total mass of the triple star system is then $M_{\rm syst} =$ $2.51^{+0.20}_{-0.18}\,M_\odot$. Additionally, the orbital parallax of the system was found to be $\pi =$ $21.9 \pm 0.6\,{\rm mas}$. A comparison with the literature results is presented in \refsec{sec:parallax}.

\subsection{Orbit of the Bab sub-system}\label{sec:orbit_Bab}

To demonstrate the capacity of their new algorithm, TRIMOR, \cite{2009MNRAS.399..906M} re-analysed the spectra of HD~188753 obtained by \cite{2007A&A...466.1179E}. As a result, they derived the radial velocities of the three stars, allowing them to classify the close pair as a double-lined spectroscopic binary (SB2). We then applied our Bayesian analysis to the RV measurements provided by \cite{2009MNRAS.399..906M} in their Table~1.

\begin{figure}[tbp]
\centering
\includegraphics[trim = 0.5cm 1.cm 0.5cm 1.cm, clip,width=0.37\textwidth ,angle=90]{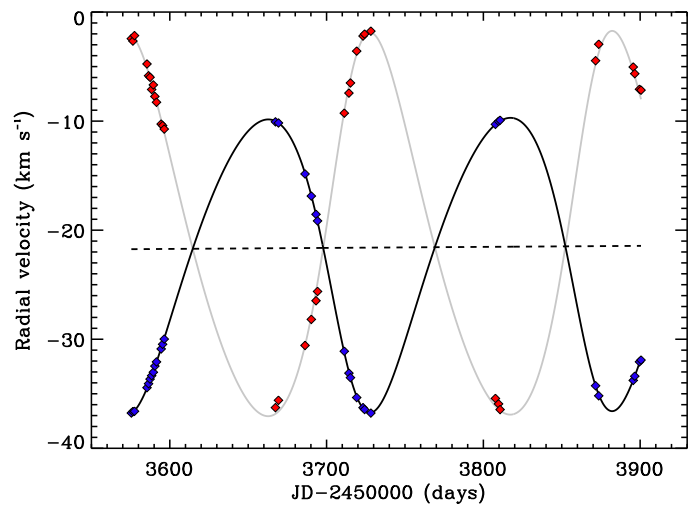}
% for a referee version
%\includegraphics[trim = 0.5cm 1.cm 0.5cm 1.cm, clip,width=0.55\textwidth ,angle=90]{RV_6469154_time.ps}
\caption{RV measurements of the Ba (blue) and Bb (red) components from \cite{2009MNRAS.399..906M}. The dashed line denotes the long-period orbital motion of the close pair and corresponds to a linear drift towards the positive radial velocities, as can be seen in the right panel of \reffig{fig:RV_6469154}. The black line denotes the RV solution of the Ba component, that is, the most massive star of the close pair, while the grey line denotes the RV solution of the Bb component, the faintest companion.}
\label{fig:RV_Mazeh}
\end{figure}

Figure~\ref{fig:RV_Mazeh} shows the best-fit solution of the radial velocities for both components of the Bab sub-system. We note that in our analysis, we included a linear drift corresponding to the long-period orbital motion of the Bab sub-system. The linear drift and the orbital parameters of the close pair, determined from our best-fit solution, are given in \reftab{tab:mazeh_B}. In particular, the Bab sub-system has an eccentricity of $0.175 \pm 0.002$ and an orbital period of $154.45 \pm 0.09\,{\rm days}$.

In the case of double-lined spectroscopic binaries, only the quantities $M_A\sin^3 i$ and $M_B\sin^3 i$ can be derived from \refeqsand{eq:massA}{eq:massB}. These quantities thus provide a lower limit of the stellar masses and a direct estimate of the mass ratio between both stars. From the orbital parameters of the close pair, presented in \reftab{tab:mazeh_B}, we obtained $M_{Ba}\sin^3 i_{Bab}= 0.258 \pm 0.004\,M_\odot$ and $M_{Bb}\sin^3 i_{Bab}= 0.198 \pm 0.002\,M_\odot$. Here, $i_{Bab}$ denotes the inclination of the Bab sub-system relative to the plane of the sky. The mass ratio of the close pair is then found to be $M_{Bb}/M_{Ba}  = 0.767 \pm 0.006$, in agreement with the value of $0.768 \pm 0.004$ provided by \cite{2009MNRAS.399..906M}. We note that our error estimate of the mass ratio is somewhat larger than that of \cite{2009MNRAS.399..906M}. We suspect that the authors underestimate the uncertainties on their orbital parameters, used in the calculation of the mass ratio, in comparison with our derived values and those of \cite{2007A&A...466.1179E}. The advantage of our Bayesian approach is that it yields credible intervals at $16\%$ and $84\%$, corresponding to the frequentist 1-$\sigma$ confidence intervals. In the case of HD~188753, the inclination of the close pair can be estimated from its spectroscopic mass sum $M_B\sin^3 i_{Bab}= 0.456 \pm 0.006\, M_{\odot}$. Indeed, the total mass of the Bab~sub-system, $M_B=1.46^{+0.13}_{-0.11}\,M_\odot$, is known from the orbital analysis of the visual pair (see \refsec{sec:orbAB}). The inclination of the close pair is then found to be $i_{Bab}= 42.8^\circ \pm 1.5^\circ$. In addition, we previously determined the inclination of the AB~system, listed in \reftab{tab:orbAB}, from our combined analysis of the visual orbit. This corresponds to a minimal relative inclination (MRI) between the two orbits of $11.7^\circ \pm 1.7^\circ$. The implications of the MRI for such a triple star system will be discussed in \refsec{sec:inclination}. 

Finally, the inclination $i_{Bab}$ can be injected in the quantities $M_{Ba}\sin^3 i_{Bab} =$ $0.258 \pm 0.004\,M_\odot$ and $M_{Bb}\sin^3 i_{Bab} =$ $0.198 \pm 0.002\,M_\odot$ in order to derive the individual masses for both stars of the close pair. We found the orbital masses for stars Ba and Bb to be $M_{Ba} =$ $0.83 \pm 0.07\,M_\odot$ and $M_{Bb} =$ $0.63^{+0.06}_{-0.05}\,M_\odot$, respectively. In the following, we will compare the orbital masses determined for the three stars of the system with the results of the stellar modelling.

%%%%%%%%%%%%%%%%%%%%%%%%%%%%%%%%%%%%%%%%%%%%%%%%%%%%%%%%%%%%%
%%%%%%%%%%%%%%%%%%%%%%%%%%Seismic data analysis%%%%%%%%%%%%%%%%%%%%%%%%%%%
%%%%%%%%%%%%%%%%%%%%%%%%%%%%%%%%%%%%%%%%%%%%%%%%%%%%%%%%%%%%%

\section{Seismic data analysis}\label{sec:seismic}

The goal of this section is to derive accurate mode frequencies and reliable proxies of the stellar masses and ages for the two seismic components of HD~188753. These quantities will then be used for the detailed modelling of each of the two components as explained in \refsec{sec:mod_opt}.

%%%%%%%%%%%%%%%%%%%%%%%%%%Mode parameter extraction%%%%%%%%%%%%%%%%%%%%%%%%%%%

\subsection{Mode parameter extraction}

The extraction of accurate mode frequencies for the two seismic components of HD~188753 is an important step before the stellar modelling. In this work, we adopted a maximum likelihood estimation approach (MLE; \citealt{1990ApJ...364..699A}) following the procedure described in \cite{2012A&A...543A..54A}, which has been extensively used during the nominal \kep{} mission to provide the mode parameters of a large number of stars \citep{2012A&A...537A.134A,2012A&A...543A..54A,2014A&A...566A..20A,2015A&A...582A..25A}.

We repeat here the different steps of this well-suited procedure for completeness:
\begin{enumerate}
\item  We derive initial guesses of the mode frequencies for the fitting procedure by applying an automated method of detection (see \citealt{2011MNRAS.415.3539V}, and references therein) which is based on the values of the seismic parameters, $\nu_{\rm max}$ and $\Delta\nu$, that are manually tweaked if required.
\item \label{step:stel-bg} We fit the power spectrum as the sum of a stellar background made up of a combination of a Lorentzian profile and white noise, as well as a Gaussian oscillation mode envelope with three parameters (the frequency of the maximum mode power, the maximum power, and the width of the mode power).
\item \label{step:fit-ps} We fit the power spectrum with $n$ orders using the mode profile model described in \cite{2015A&A...582A..25A}, with no rotational splitting and the stellar background fixed as determined in step \ref{step:stel-bg}.
\item We repeat step \ref{step:fit-ps} but leave the rotational splitting and the stellar inclination angle as free parameters, and then apply a likelihood ratio test to assess the significance of the fitted splitting and inclination angle.
\end{enumerate}
The steps above were used for the main mode power at $2200\,\mu{\rm Hz}$, and repeated for the mode power at $3300\,\mu{\rm Hz}$. We note that for the secondary component, we used the residual power spectrum derived from the fit of the primary. It corresponds to the ratio between the observed power spectrum and the best fitting model of the primary component, which includes a Lorentzian profile for the stellar background. As a result, the background was modelled with a single white noise component when fitting the mode power at $3300\,\mu{\rm Hz}$. The procedure for the quality assurance of the frequencies obtained for both stars is described in \cite{2012A&A...543A..54A} with a slight modification (for more details, see \citealt {2015A&A...582A..25A}). The frequencies and their formal uncertainties, derived from the inverse of the Hessian matrix, are provided in \refapp{app:osc_freq}.

%%%%%%%%%%%%%%%%%%%%%%%%%%Stellar parameters from scaling relations%%%%%%%%%%%%%%%%%%%%%%%%%%%

\subsection{Stellar parameters from scaling relations}\label{sec:scal_rel}
Using the seismic parameters given in \reftab{tab:stellar_param} for stars A and Ba, it is possible to determine their stellar parameters without further advanced modelling. Indeed, the stellar mass and radius of the two seismic components can be estimated from the well-known scaling relations:
\begin{equation}
\frac{M}{M_\odot} = \left(\frac{\nu_{{\rm max}}}{\nu_{{\rm ref}}}\right)^3 \left(\frac{\Delta\nu}{\Delta\nu_{{\rm ref}}}\right)^{-4} \left(\frac{T_{{\rm eff}}}{T_{{\rm eff,}\odot}}\right)^{3/2},\label{eq:m_scal}
\end{equation}
\begin{equation}
\frac{R}{R_\odot} = \left(\frac{\nu_{{\rm max}}}{\nu_{{\rm ref}}}\right) \left(\frac{\Delta\nu}{\Delta\nu_{{\rm ref}}}\right)^{-2} \left(\frac{T_{{\rm eff}}}{T_{{\rm eff,}\odot}}\right)^{1/2},\label{eq:r_scal}
\end{equation}
where $\nu_{\rm max}$ is the frequency of maximum oscillation power, $\Delta\nu$ is the large frequency separation,  and $T_{\rm eff}$ is the effective temperature of the star. Here, we adopted the reference values from \citet{Mosser2013}, $\nu_{\rm ref} = 3104\,\mu$Hz and $\Delta\nu_{\rm ref} = 138.8\,\mu$Hz, and the solar effective temperature, $T_{{\rm eff,}\odot} = 5777\,$K.

For each seismic component, we derived $\nu_{\rm max}$ by fitting a parabola over four (star Ba) or five (star A) monopole modes around the maximum of mode height and $\Delta\nu$ by fitting the asymptotic relation for frequencies \citep{1980ApJS...43..469T}. 
%Asymptotic large frequency separation ($\Delta\nu_{\rm as}$) as given by Eq. (6) of \citet{Mosser2013}.
For star A, we obtained $\nu_{\rm max} =$ $2204 \pm 8\,\mu{\rm Hz}$ and $\Delta\nu =$ $106.96 \pm 0.08\,\mu{\rm Hz}$, while for star Ba, we obtained $\nu_{\rm max} =$ $3274 \pm 67\,\mu{\rm Hz}$ and $\Delta\nu =$ $147.42 \pm 0.08\,\mu{\rm Hz}$. In this work, we adopted the spectroscopic values of the effective temperature from \cite{2005Natur.436..230K}, $T_{{\rm eff,A}} =$ $5750 \pm 100\,$K and $T_{{\rm eff,Ba}} =$ $5500 \pm 100\,$K, measured independently for stars~A and~Ba. Using \refeqsand{eq:m_scal}{eq:r_scal} with the above values, we then obtained  $M_{\rm A} =$ $1.01 \pm 0.03\,M_\odot$ and $R_{\rm A} =$ $1.19 \pm 0.01\,R_\odot$ for star~A, and $M_{\rm Ba} =$ $0.86 \pm 0.06\,M_\odot$ and $R_{\rm Ba} =$ $0.91 \pm 0.02\,R_\odot$ for star~Ba. The effective temperature of Ba is contaminated by that of Bb.  Due to the flux ratio of about 1 to 9 as measured by \cite{2009MNRAS.399..906M}, the temperature of Ba is likely to be 3\% lower. With the effective temperature bias due to Bb, the mass and the radius of Ba would be lower by 0.04$\,M_\odot$, 0.015$\,R_\odot$, respectively; still commensurate with the random errors.  We also deduced the luminosity of the two stars from the Stefan-Boltzmann law, $L \propto R^{2} T_{{\rm eff}}^{4}$, as $L_{\rm A} =$ $1.40 \pm 0.12\,L_\odot$ and $L_{\rm Ba} =$ $0.69 \pm 0.07\,L_\odot$ (with a bias of -0.1$\,L_\odot$ due to the presence of Bb).

Consequently, provided that a measurement of $T_{\rm eff}$ is available, the seismic parameters $\nu_{\rm max}$ and $\Delta\nu$ give a direct estimate of the stellar mass and radius independent of evolutionary models. This so-called direct method has been applied to the hundreds of solar-type stars for which \kep{} has detected oscillations \citep{2011Sci...332..213C,2014ApJS..210....1C}. In the case of HD~188753, the masses derived for each star from the direct method are consistent with the results of our orbital analysis. Therefore, these estimates provide an initial range of mass for stellar modelling. 

%%%%%%%%%%%%%%%%%%%%%%%%%%Frequency separation ratios%%%%%%%%%%%%%%%%%%%%%%%%%%% 

\subsection{Frequency separation ratios}
%Different combinations of oscillation frequencies have been presented in the literature for studying the internal structure of stars. In particular, \cite{Roxburgh2003a} proposed to use the ratios of small to large separations as a diagnostic of the deep stellar interiors. 

In this work, we used the frequency separation ratios as observational constraints for the model fitting, instead of the individual frequencies themselves. Indeed, \cite{Roxburgh2003a} demonstrated that the frequency ratios are approximately independent of the structure of the outer layers and are determined only by the internal structure of the star. They are therefore less sensitive than the individual frequencies to the improper modelling of the near-surface layers (\citealt{2005A&A...434..665R,Floranes2005}) and are also insensitive to the line-of-sight Doppler velocity shifts \citep{Davies2014}. The use of these ratios then allows us to avoid potential biases from applying frequency corrections for near-surface effects (\eg \citealt{2008ApJ...683L.175K,2013MNRAS.435..242G,2014A&A...568A.123B,2014A&A...569C...2B}) and for Doppler shifts.

\begin{figure*}[tbp]
\centering
\vbox{
\includegraphics[trim = 1.cm 6.5cm 2.cm 7.5cm, clip,width=0.49\textwidth ,angle=0]{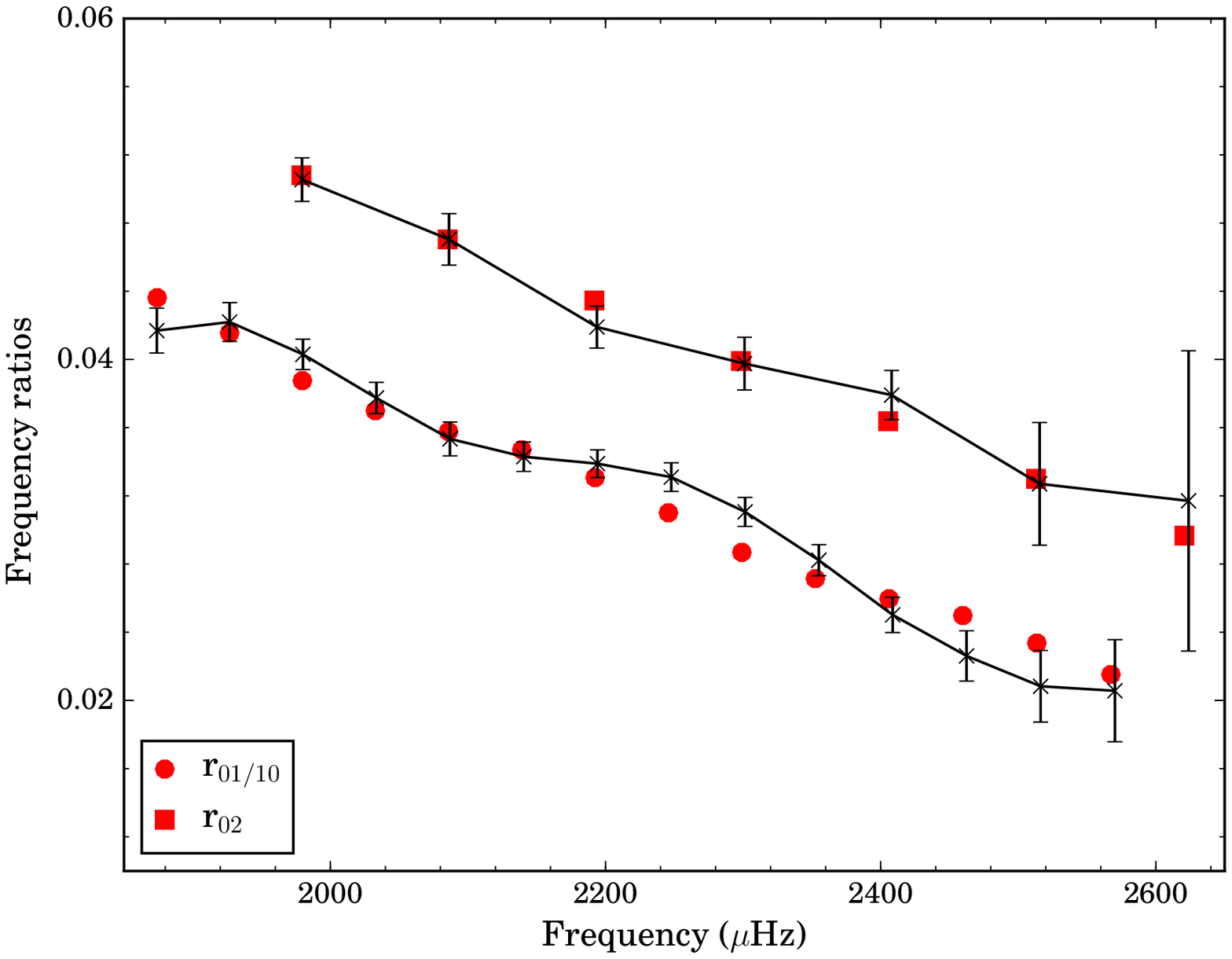}
\includegraphics[trim = 1.cm 6.5cm 2.cm 7.5cm, clip,width=0.49\textwidth ,angle=0]{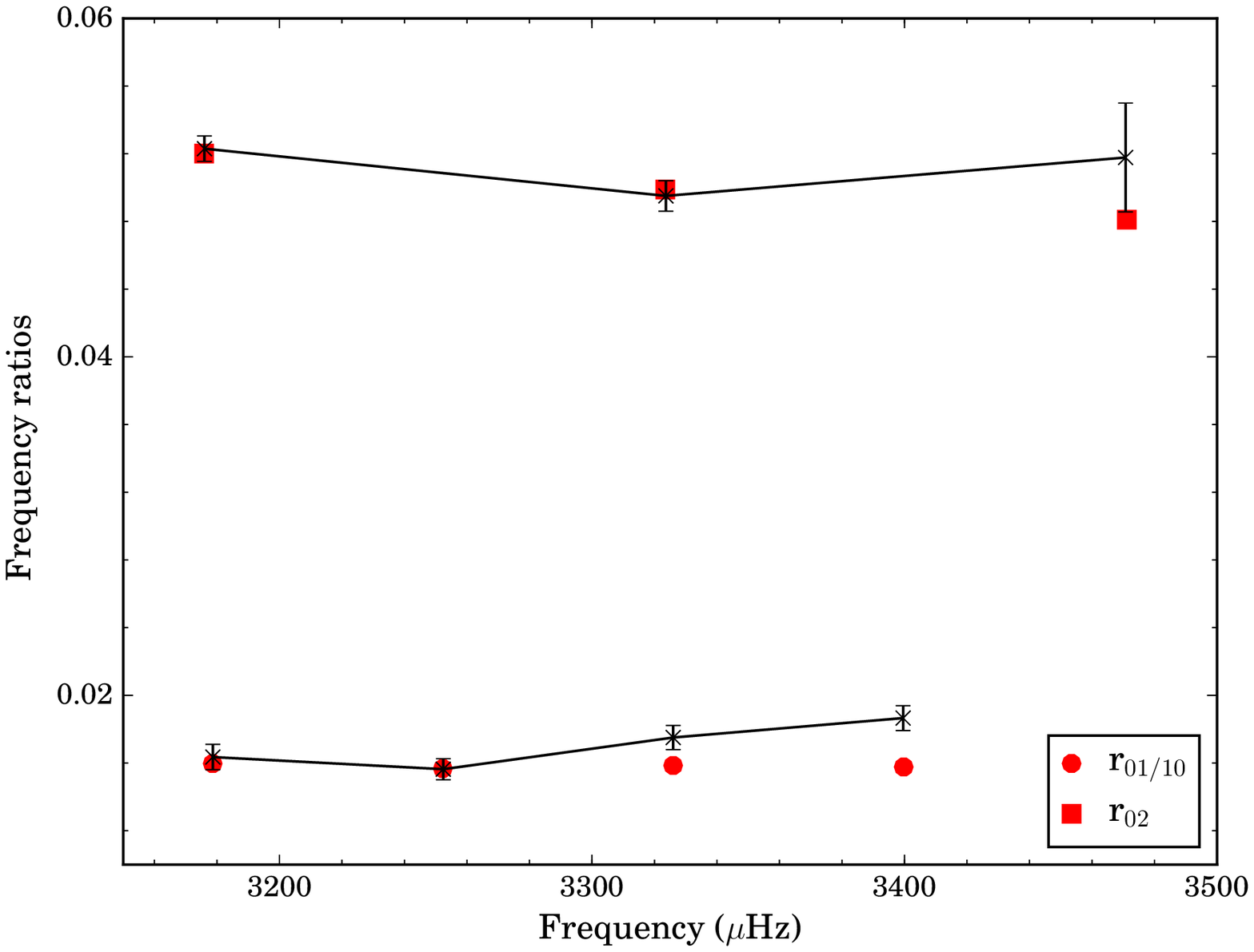}
}
\caption{Frequency separation ratios as a function of frequency for star~A (left) and star~Ba (right). The ratios $r_{01/10}(n)$ and $r_{02}(n)$ derived from the reference model of each star are indicated by red circles and squares, respectively, while the observed ratios are shown as connected points with errors.}
\label{fig:ratios}
\end{figure*}

The frequency separation ratios are defined as:
\begin{equation}
r_{02}(n) = \frac{d_{02}(n)}{\Delta_1(n)}\mbox{,}\label{eq:r02}
\end{equation}
\begin{equation}
r_{01}(n) = \frac{d_{01}(n)}{\Delta_1(n)}\mbox{,}\qquad r_{10}(n) = \frac{d_{10}(n)}{\Delta_0(n+1)},\label{eq:r01}
\end{equation}
where $d_{02}(n) =$ $\nu_{n,0} - \nu_{n-1,2}$ and $\Delta_{l}(n) =$ $\nu_{n,l} - \nu_{n-1,l}$ refer to the small and large separations, respectively. Here, we adopted the five-point smoothed small frequency separations following \cite{Roxburgh2003a}:
\begin{equation}
 d_{01}(n) = \frac{1}{8}(\nu_{n-1,0} - 4\nu_{n-1,1} + 6\nu_{n,0} - 4\nu_{n,1} +\nu_{n+1,0}),
\end{equation}
\begin{equation}
d_{10}(n) =  - \frac{1}{8}(\nu_{n-1,1} - 4\nu_{n,0} + 6\nu_{n,1} - 4\nu_{n+1,0} +\nu_{n+1,1}),
\end{equation}
where $\nu$ is the mode frequency, $n$ is the radial order and $l$ is the angular degree. In \reffig{fig:ratios}, we plotted for both stars the ratios $r_{01}$, $r_{10}$ and $r_{02}$ computed from the observed frequencies. A clear oscillatory behaviour of the observed frequency ratios can be seen for star~A, in the left panel of \reffig{fig:ratios}, indicating an acoustic glitch. Such a signature arises from the acoustic structure and has already been detected in other solar-like stars \citep{Mazumdar2014}.

Another advantage of using the frequency ratios is that the ratio $r_{02}$ is sensitive to the structure of the core and hence to the age of the star (\citealt{Floranes2005,Lebreton2009}). In particular, \cite{2011ApJ...743..161W} used the ratio $r_{02}$ instead of the small frequency separation to construct a modified version of the so-called C-D diagram \citep{JCD84b}, showing that this ratio is an effective indicator of the stellar age. In the case of HD~188753, the mean values of the measured ratios, averaged over the whole range of the observed radial orders, are $\langle r_{02} \rangle_A =$ $0.043 \pm 0.001$ and $\langle r_{02} \rangle_{Ba} =$ $0.051 \pm 0.001$. From these values and those of the large frequency separation listed in \reftab{tab:stellar_param}, we can then determine the position of the two seismic components in the modified C-D diagram obtained by \cite{2011ApJ...743..161W}. Using their Fig.~8, we found an age of $\sim$11$\,{\rm Gyr}$ for star A and of $\sim$8.0$\,{\rm Gyr}$ for star Ba. 

In addition, \cite{2017A&A...601A..67C} recently derived a linear relation between the mean value of $r_{02}$ and the stellar age from the analysis of 57~stars observed by \kep{}. Using their Eq.~(6) with the above values of $\langle r_{02} \rangle$, we found the individual ages for stars~A and~Ba to be $9.5 \pm 0.2\,{\rm Gyr}$ and $8.0 \pm 0.2\,{\rm Gyr}$, respectively. We point out that our estimated errors only result from the propagation of the uncertainties on the measured ratios, since \cite{2017A&A...601A..67C} do not provide uncertainties on their fitted parameters. The use of the ratio $r_{02}$ as a proxy of the stellar age therefore allows us to derive an initial range of age for stellar modelling.

%%%%%%%%%%%%%%%%%%%%%%%%%%%%%%%%%%%%%%%%%%%%%%%%%%%%%%%%%%%%%
%%%%%%%%%%%%%%%%%%%%%%%%%%Stellar models%%%%%%%%%%%%%%%%%%%%%%%%%%%
%%%%%%%%%%%%%%%%%%%%%%%%%%%%%%%%%%%%%%%%%%%%%%%%%%%%%%%%%%%%%

\section{Stellar modelling}\label{sec:modelling}

%%%%%%%%%%%%%%%%%%%%%%%%%%Input physics%%%%%%%%%%%%%%%%%%%%%%%%%%% 
 
\subsection{Input physics}\label{sec:input_phys}

In order to determine precise stellar parameters for HD~188753, we performed a detailed modelling of each seismic component using the stellar evolution code CESTAM (Code d'\'{E}volution Stellaire, avec Transport, Adaptatif et Modulaire; \citealt{2008Ap&SS.316...61M,2013A&A...549A..74M}).

For our model calculations, we adopted the 2005 version of the OPAL equation of state \citep{2002ApJ...576.1064R} and the OPAL opacities \citep{1996ApJ...464..943I} complemented by those of \cite{1994ApJ...437..879A} for low temperatures. The opacity tables are given for the new solar mixture derived by \cite{2009ARA&A..47..481A}, which corresponds to $(Z/X)_\odot = 0.0181$. The microscopic diffusion of helium and heavy elements, including gravitational settling, thermal and concentration diffusion, but no radiative levitation, was taken into account following the prescription of \cite{1993ASPC...40..246M}. We used the NACRE nuclear reaction rates \citep{ang99} with the revised ${}^{14}\text{N(p},\gamma)^{15}\text{O}$ reaction rate from \cite{Formicola2004}. Convection was treated according to the mixing-length theory (MLT; \citealt{1958ZA.....46..108B}) and no overshooting was considered. A standard Eddington gray atmosphere was employed for the atmospheric boundary condition.

%%%%%%%%%%%%%%%%%%%%%%%%%%Model optimisation%%%%%%%%%%%%%%%%%%%%%%%%%%% 

\subsection{Model optimisation}\label{sec:mod_opt}

For each seismic component of HD~188753, we constructed a grid of stellar evolutionary models using CESTAM with the input physics described above. The goodness of fit was evaluated for all of the grid models from the merit function $\chi^2$ given by:
\begin{equation}
\chi^2 = (x_{{\rm mod}}-x_{{\rm obs}})^{{\rm T}} \, {\rm C}^{-1} \,  (x_{{\rm mod}}-x_{{\rm obs}}),\label{eq-chi2}
\end{equation}
where $x_{\rm obs}$ denotes the observational constraints considered, that is, the frequency ratios defined in \refeqsand{eq:r02}{eq:r01}, where $x_{\rm mod}$ denotes the theoretical values predicted by the model and T denotes the transposed matrix. Here, C refers to the covariance matrix of the observational constraints, which is not diagonal due to the strong correlations between the frequency ratios. The theoretical oscillation frequencies used for the calculation of the frequency ratios were computed with the Aarhus adiabatic oscillation package (ADIPLS; \citealt{jcd08adipls}).

Firstly, a total of 2000 random grid points were drawn for each star from a reasonable range of the stellar parameters, these being, the age of the star, the mass $M$, the initial helium abundance $Y_0$, the initial metal-to-hydrogen ratio $(Z/X)_0$ and the mixing-length parameter for convection $\alpha_{\rm MLT}$. Following the age-$\langle r_{02} \rangle$ relation of \cite{2017A&A...601A..67C}, the age of the two stars was assumed to be in the range 5--13$\,{\rm Gyr}$. For stars~A and~Ba, we adopted $M_A \in$ $[0.9,1.2]\,M_\odot$ and $M_{Ba} \in$ $[0.7,1.0]\,M_\odot$, respectively, in agreement with the results of the orbital analysis and scaling relations. We considered $Y_0 \in$ $[Y_P,0.31]$ where $Y_P =$ $0.2477 \pm 0.0001$ is the primordial value from standard Big Bang nucleosynthesis \citep{2014A&A...571A..16P}. The present $(Z/X)$ ratio can be derived from the observed [Fe/H] value, provided in \reftab{tab:stellar_param}, through the relation $[{\rm Fe}/{\rm H}] = \log(Z/X) - \log(Z/X) _\odot$. We then estimated the initial $[Z/X]_0$ ratio as being in the range 0.01--0.05, which was shifted to a higher value to compensate for diffusion. We adopted $\alpha_{\rm MLT} \in$ $[1.6,2]$ following the solar calibration of \cite{2014MNRAS.445.4366T} that corresponds to $\alpha_{{\rm MLT,}\odot} = 1.76 \pm 0.04$.

Secondly, in order to determine the final parameters of the two stars, we employed a Levenberg-Marquardt minimisation method using the Optimal Stellar Models (OSM)\footnote{\url{https://pypi.python.org/pypi/osm}} software developed by R.~Samadi.
%Optimal Stellar Models (OSM)\footnote{OSM software: \url{https://pypi.python.org/pypi/osm}}
For star~A, we selected the 50 grid models with the lowest $\chi^2$ values and used them as a starting point for the final optimisation. We then performed a set of local minimisations adopting the age of the star, the mass, the initial helium abundance, the initial $(Z/X)_0$ ratio and the mixing-length parameter as free model parameters. Again, we adopted the merit function defined in \refeq{eq-chi2} associated with the frequency ratios. For star~Ba, due to the limited number of observed ratios, we decided to only adjust the age, the mass, and the initial helium abundance. We then performed a first set of minimisations from the 15 grid models with the lowest $\chi^2$ values. During the minimisation process, the initial $(Z/X)_0$ ratio and the mixing-length parameter were fixed at the value of the grid point considered. In order to explore the impact of changing these parameters, we performed a second set of minimisations for the 15 grid models previously selected adopting $(Z/X)_0 = $ $[0.015,0.025,0.035,0.045]$ and $\alpha_{\rm MLT} =$ $[1.65,1.75,1.85,1.95]$ as fixed values. In total, the number of local minimisations for star~Ba was $15 + 4 \times 4 \times 15 =$ $255$. We note that our reference model of star~Ba, provided in \reftab{tab:seismic_model}, was obtained for $(Z/X)_0$ and $\alpha_{\rm MLT}$ fixed at their grid-point values.

\reffig{fig:ratios} shows the comparison of the observed frequency ratios (black crosses) with the corresponding values derived from the reference model of each star (red symbols). The modelling results for stars~A and~Ba are given in \reftab{tab:seismic_model}. The error bars on the stellar parameters were derived from the inverse of the Hessian matrix. The normalised $\chi^2$ values were computed as $\chi^2_N =$ $\chi^2/N$ where $N$ denotes the number of observed frequency ratios used for the optimisation.

\begin{table}[tbp]
\caption{Reference models for star A and star Ba.}
\centering
\begin{tabular}{l c c} 
\hline
\hline
Parameter &  Star A & Star Ba \\
\hline
$M$ ($M_\odot$) & $0.99 \pm 0.01$ & $0.86 \pm 0.01$ \\
Age (Gyr) & $10.7 \pm 0.2$ & $11.0 \pm 0.3$ \\
$Y_0$ & $0.278 \pm 0.002$ & $0.289 \pm 0.004$ \\
$(Z/X)_0$ & $0.033 \pm 0.001$ & 0.031\tablefootmark{a}\\
$\alpha_{\rm MLT}$ & $1.83 \pm 0.06$ & 1.77\tablefootmark{a}\\
\hline
$L$ ($L_\odot$) & 1.23 & 0.57 \\
$R$ ($R_\odot$) & 1.18 & 0.91 \\
$T_{\rm eff}$ (K) & 5598 & 5272 \\
$\log g$ (cgs) & 4.29 & 4.46 \\
${\rm [Fe/H]}$ (dex) & 0.17 & 0.15 \\
%{\red $\chi^2$} & {\red 49.3574\tablefootmark{b}} & {\red 29.1567\tablefootmark{b}} \\
$\chi^2_N$ & 2.35 & 4.17  \vspace*{0.05cm} \\ 
\hline
\end{tabular}
\tablefoot{%
\tablefoottext{a}{The reference model of star~Ba was obtained for $(Z/X)_0$ and $\alpha_{\rm MLT}$ fixed at their grid-point values during the minimisation process (see text).}
%\tablefoottext{b}{{\red[temp]}}
}
\label{tab:seismic_model}
\end{table}

%%%%%%%%%%%%%%%%%%%%%%%%%%%%%%%%%%%%%%%%%%%%%%%%%%%%%%%%%%%%%
%%%%%%%%%%%%%%%%%%%%%%%%%%Results and discussion%%%%%%%%%%%%%%%%%%%%%%%%%%%
%%%%%%%%%%%%%%%%%%%%%%%%%%%%%%%%%%%%%%%%%%%%%%%%%%%%%%%%%%%%%

\section{Results and discussion}\label{sec:results}

%%%%%%%%%%%%%%%%%%%%%%%%%%Physical parameters of HD 188753%%%%%%%%%%%%%%%%%%%%%%%%%%%

\subsection{Physical parameters of HD~188753}\label{sec:phys-param}

From the modelling, we found the individual ages for stars~A and~Ba to be $10.7 \pm 0.2\,{\rm Gyr}$ and $11.0 \pm 0.3\,{\rm Gyr}$, respectively. These values are consistent within their error bars, as expected from the common origin of both stars. Combining the individual ages, the system is then about $10.8 \pm 0.2\,{\rm Gyr}$ old. 
 
In addition, the stellar masses from our reference models were found to be $M_A =$ $0.99 \pm 0.01\,M_\odot$ and $M_{Ba} =$ $0.86 \pm 0.01\,M_\odot$, which agree very well with the results of the orbital analysis and scaling relations. This latter value can be injected in the mass ratio of the close pair derived in \refsec{sec:orbit_Bab}, $M_{Bb}/M_{Ba}  =$ $0.767 \pm 0.006$, in order to determine the mass of the faintest star with much more precision than in our orbital analysis. The seismic mass of the star~Bb is then found to be $M_{Bb} =$ $0.66 \pm 0.01\,M_\odot$. As a result, the seismic mass of the Bab sub-system is $M_B = 1.52 \pm 0.02\,M_\odot$ and the total seismic mass of the triple star system is $M_{\rm syst} =$ $2.51 \pm 0.02\,M_\odot$. This method provides a precise but model-dependent total mass of the system, which can be compared with the results of our orbital analysis. Indeed, as explained in \refsec{sec:orbAB}, we obtained a direct estimate of the total mass, $M_{\rm syst} =$ $2.51^{+0.20}_{-0.18}\,M_\odot$, using the astrometric and radial-velocity observations of the system. We find excellent agreement between the seismic and orbital values of the total mass for HD~188753, bolstering our confidence in the results of the stellar modelling.

Our stellar models thus provide precise and accurate values of the individual masses that can be used to refine the physical parameters of the system. For example, the semi-major axis of the relative orbit is related to the total mass and to the orbital period of the system from the Kepler's third law. In other words, the total mass of the system expressed in units of the solar mass is given by $M_{\rm syst} =$ $a_{\rm AU}^3 /P^2$, where $a_{\rm AU}$ is the semi-major axis expressed in astronomical units and $P$ is the orbital period expressed in years. Using the Kepler's third law with the total seismic mass and the orbital period of the AB~system, $M_{\rm syst} =$ $2.51 \pm 0.02\,M_\odot$ and $P =$ $25.63 \pm 0.04\,{\rm years}$, we find that the semi-major axis of the relative orbit for the visual pair is $a_{\rm AU} =$ $11.82 \pm 0.03\,{\rm AU}$. Furthermore, by applying the definition of the centre of mass to the AB system, we can also derive the semi-major axis of the two barycentric orbits with:
\begin{equation}
a_A = a_{\rm AU}\,\frac{M_B}{M_A+M_B}\quad {\rm and} \quad a_B = a_{\rm AU}\,\frac{M_A}{M_A+M_B},
\end{equation}
where, obviously, $a_{\rm AU} =$ $a_A+a_B$. Adopting the seismic masses $M_A =$ $0.99 \pm 0.01\,M_\odot$ and $M_B = 1.52 \pm 0.02\,M_\odot$ and the above value of the semi-major axis $a_{\rm AU}$, we then obtain $a_{A} =$ $7.17 \pm 0.04\,{\rm AU}$ and $a_{B} =$ $4.65 \pm 0.03\,{\rm AU}$. Similarly, the Kepler's third law and the definition of the centre of mass can be applied to the Bab sub-system using the seismic masses of the close pair. From the above values of $M_{Ba}$ and $M_{Bb}$, associated with an orbital period of $154.45 \pm 0.09\,{\rm days}$, we find that the semi-major axes for star~Ba and star~Bb are $a_{Ba} =$ $0.282 \pm 0.002\,{\rm AU}$ and $a_{Bb} =$ $0.367 \pm 0.002\,{\rm AU}$, respectively.

%%%%%%%%%%%%%%%%%%%%%%%%%%Asteroseismic parallax%%%%%%%%%%%%%%%%%%%%%%%%%%%

\subsection{Asteroseismic parallax}\label{sec:parallax}

As explained in \refsec{sec:orbAB}, the astrometric and radial-velocity observations of the visual pair can be combined in order to derive the parallax of the system. The term `orbital parallax' has been suggested to denote a parallax determined in this way \citep{1992AJ....104..241A}. From our combined analysis of the visual pair, we then obtained an orbital parallax of $\pi = 21.9 \pm 0.6\,{\rm mas}$.

In the case of HD~188753, the precision on the parallax can be improved using the total seismic mass of the system determined in \refsec{sec:phys-param}. Indeed, combining the Kepler's third law with \refeq{eq:parallax} provides a direct relation between the total mass and the parallax of the system:\\
\begin{equation}
M_{\rm syst} = \left(\frac{a_{\rm mas}}{\pi_{\rm mas}}\right)^3\frac{1}{P^2},\label{eq:msyst}
\end{equation}
where $M_{\rm syst}$ is expressed in units of the solar mass and $\pi_{\rm mas}$ is expressed in milliarcseconds. Here, $a_{\rm mas}$ is the angular semi-major axis of the visual pair, in mas, and $P$ is the orbital period, in years. Adopting the total seismic mass $M_{\rm syst} =$ $2.51 \pm 0.02\,M_\odot$ and the values of $a_{\rm mas}$ and $P$ listed in \reftab{tab:orbAB}, we find that the `asteroseismic parallax' of the system is $\pi = 21.9 \pm 0.2\,{\rm mas}$. The error bars on the parallax are estimated through a Monte Carlo simulation using the chains from our Bayesian analysis for $a_{\rm mas}$ and $P$, and a randomised total mass. The probability distributions of the three parameters are then injected in \refeq{eq:msyst} in order to compute the median and credible intervals for the parallax. We thus obtain a precise estimate of the parallax by combining the asteroseismic and astrometric observations of the system.
 
As a comparison, the literature values of the parallax for HD~188753 are given in \reftab{tab:parallax}. Firstly, we note that our results are consistent with the revised Hipparcos parallax, $\pi =$ $21.6 \pm 0.7\,{\rm mas}$, obtained by \cite{2007AA...474..653V}. For HD~188753, the author adopted a standard model characterised by five astrometric parameters to describe the apparent motion of the source. These five parameters are the position $(\alpha,\delta)$ at the Hipparcos epoch, the parallax $\pi$ and the proper motion $(\mu_{\alpha *},\mu_\delta)$, and refer to the photocenter of the system. In the case of resolved binaries, it is possible to take the duplicity of the system into account by combining the Hipparcos astrometry with existing ground-based observations. For this, \cite{1999AA...341..121S} employed an astrometric model specified by the seven orbital parameters $(P,T,e,\omega,a,i,\Omega)$, also known as the Campbell elements, in addition to the five parameters quoted above. For HD~188753, \cite{1999AA...341..121S} then obtained a parallax of $21.9 \pm 0.6\,{\rm mas}$, which is in excellent agreement with the results of our analysis. Using stellar modelling, we thus reduced by a factor of about three the uncertainty on the parallax, which is found independently of the Hipparcos data. In addition, the combined treatment of the visual pair, using (visual and speckle) relative positions plus Hipparcos data, allows the author to derive the semi-major axis and the orbital period defined in \refeq{eq:msyst}. Knowing the parallax, \cite{1999AA...341..121S} found that the total mass of the system is $M_{\rm syst} =$ $2.73 \pm 0.33\,M_\odot$. The discrepancy between this latter estimate and that reported in this work can be explained by the larger value of the semi-major axis derived by \cite{1999AA...341..121S}. However, we point out that our orbital solution results from the derivation of a full orbit using high-resolution techniques, in contrast to that of \cite{1999AA...341..121S}.

\begin{table}[tbp]
\caption{Estimated values of the parallax for HD~188753.}
\centering
\begin{tabular}{c c} 
\hline
\hline
Parallax & Reference \\
(mas) &  \\
\hline
%$21.9 \pm 0.2$ & in this article\tablefootmark{a} \\
%$21.9 \pm 0.6$ & in this article\tablefootmark{b} \\
$21.9 \pm 0.2$ & our seismic estimate\tablefootmark{a} \\
$21.9 \pm 0.6$ & our orbital estimate\tablefootmark{b} \\
%$20.1 \pm 0.4$ & Gaia website\tablefootmark{c}  \\
$21.6 \pm 0.7$ & \cite{2007AA...474..653V}  \\
$21.9 \pm 0.6$  & \cite{1999AA...341..121S}  \\
\hline
\end{tabular}
\tablefoot{%
\tablefoottext{a}{See \refsec{sec:parallax}.}
\tablefoottext{b}{See \refsec{sec:orbAB}.}%\\
%\tablefoottext{a}{\url{https://gea.esac.esa.int/archive/}}
%\tablefoottext{c}{\url{https://gea.esac.esa.int/archive/}}
%\tablefoottext{c}{Gaia Archive: \url{https://gea.esac.esa.int/archive/}}
}
\label{tab:parallax}
\end{table}

%Secondly, we also added in \reftab{tab:parallax} the most recent value of the parallax for HD~188753 from the Gaia Data Release~1 (DR1; \citealt{2016A&A...595A...4L}). We note a significant disagreement between the Gaia DR1 parallax, $\pi =$ $20.1 \pm 0.4\,{\rm mas}$, and the other values presented in this work. 
Unfortunately, there is no available parallax for the system HD~188753 from the Gaia Data Release~1 (DR1; \citealt{2016A&A...595A...4L}). However, as explained in \cite{2016A&A...595A...4L}, binaries and multiple stellar systems did not receive a special treatment in the data processing. For Gaia DR1, the sources were all treated as single stars. Furthermore, the derived proper motion should be interpreted as the mean motion of the system between the Hipparcos epoch (J1991.25) and the Gaia DR1 epoch (J2015.0). The consequence could be that the positions at the two epochs refer to different components of the system, including its photocentre.
%Resolved system: $0.1 \lesssim  \rho  \lesssim 10\,{\rm arcsec}$ and $\Delta V \lesssim 4\,{\rm mag}$.\\ 

%%%%%%%%%%%%%%%%%%%%%%%%%%Relative inclination of the two orbits%%%%%%%%%%%%%%%%%%%%%%%%%%%

\subsection{Relative inclination of the two orbits}\label{sec:inclination}

From our orbital analysis of HD~188753, we derived in \refsec{sec:orbit_Bab} a minimal relative inclination (MRI) between the visual AB~system and the close Bab~sub-system of $11.7^\circ \pm 1.7^\circ$. In particular, we used the orbital value of the mass $M_B$, associated with the quantity $M_B\sin^3 i_{Bab}= 0.456 \pm 0.006\, M_{\odot}$, to estimate the inclination of the Bab~sub-system. 

As demonstrated in \refsec{sec:phys-param}, stellar modelling provides precise and accurate values of the individual masses, which can be adopted in our calculations to refine the physical parameters of the system. Using the above value of $M_B\sin^3 i_{Bab}$ with the seismic mass $M_B =$ $1.52 \pm 0.02\,M_\odot$, we find that the inclination of the Bab~sub-system is $i_{Bab} =$ $42.0^\circ \pm 0.3^\circ$. Since the inclination of the AB~system is known (see \reftab{tab:orbAB}), the minimal relative inclination between the two orbits can be precisely determined. We thus obtain a final MRI value of $10.9^\circ \pm 1.5^\circ$ for the triple star system HD~188753. \cite{2009MNRAS.399..906M} attempted to estimate the relative inclination of the system using a similar approach. For this, the authors adopted the total mass of the system provided by \cite{1999AA...341..121S} and the mass ratio of the visual pair from \cite{2007A&A...466.1179E} to determine the mass of the Bab~sub-system. The derived value was then injected in Eq.~(6) of \cite{2009MNRAS.399..906M}, which is equivalent to the spectroscopic mass sum $M_B\sin^3 i_{Bab}$ of the close pair. In this way, \cite{2009MNRAS.399..906M} found that the inclination of the Bab~sub-system is $i_{Bab} =$ $39.6^\circ \pm 2.8^\circ$ (their Table~3). Considering an inclination for the visual pair of 34$^\circ$ \citep{1999AA...341..121S}\footnote{\cite{1999AA...341..121S} value is not accompanied by error bars.}, the MRI value obtained by \cite{2009MNRAS.399..906M} from their astrometric approach is about~6$^\circ$. The advantage of our approach is to provide a more robust estimate of the MRI by combining the modelling results with those derived from the orbital analysis.

The distribution of the relative inclination in triple stars can provide valuable information about the dynamical evolution of multiple stellar systems \citep{2002A&A...384.1030S}. The relative inclination, $\phi$, between the two orbital planes is given by \citep{1973bmss.book.....B,1981ApJ...246..879F}:
\begin{equation}
\cos\phi= \cos i_{\rm out} \cos i_{\rm in}+ \sin i_{\rm out} \sin i_{\rm in} \cos(\Omega_{\rm out}-\Omega_{\rm in}),\label{eq:phi}
\end{equation}
where $i$ is the orbital inclination, $\Omega$ is the position angle of the ascending node and $\phi \in [0,\pi]$. Here, the indices `out' and `in' refer to the outer and inner orbits, namely the AB~system and the Bab sub-system, respectively. We note that the inclination $\phi$ also corresponds to the relative angle between the angular momentum vectors of the inner and outer orbits. As a result, the mutual orientation of both orbits is prograde for $0^\circ \leq \phi < 90^\circ$ and retrograde for $90^\circ < \phi \leq 180^\circ$. In the case of HD~188753, the position angle of the ascending node is totally unknown for the Bab~sub-system and thus the relative inclination $\phi$ cannot be determined. However,
it is easy to show from \refeq{eq:phi} that $i_{Bab} - i_{AB} \leq \phi \leq i_{Bab} + i_{AB}$. From the derived values of $i_{Bab}$ and $i_{AB}$, we then obtain a relative inclination of $10.9^\circ \leq \phi_{\rm pro} \leq 73.1^\circ$ where the lower limit corresponds to the MRI defined above. Since the direction of motion is not known from radial-velocity observations, we point out that the orbital inclination of the Bab~sub-system may also be $i_{Bab} =$ $138.0^\circ \pm 0.3^\circ$. By symmetry, we then find that $106.9^\circ \leq \phi_{\rm retro} \leq 169.1^\circ$.

In contrast to the previous analysis of \cite{2009MNRAS.399..906M}, we argue that the triple star system HD~188753 does not have a coplanar configuration. Indeed, we found that the relative inclination between the two orbits may be as high as $\sim$73$^\circ$, with a lower limit at $10.9^\circ \pm 1.5^\circ$. For inclined systems, it has been shown that the eccentricity of the inner orbit, $e_{\rm in}$, and the relative inclination, $\phi$, may vary periodically such that the quantity $(1-e_{\rm in}^2) \cos^2 \phi$ is conserved. This so-called Lidov-Kozai mechanism \citep{1962P&SS....9..719L,1962AJ.....67..591K}, which takes place when $39.2^\circ \leq \phi  \leq 140.8^\circ$, plays an important role in the evolution of multiple stellar systems (see \eg \citealt{2016ComAC...3....6T} for a review). Unfortunately, the suitability of such a mechanism cannot be assessed for our triple star system. However, it is interesting to see how the combined analysis of HD~188753 enabled us to put stringent limits on its relative inclination.

%$\phi$: relative angle between the angular momentum vectors of the inner and outer orbits.\\
%$\phi \in [0,\pi]$: relative angle between the angular momentum vectors of the AB system and the Bab sub-system, also known as obliquity.\\
%Obliquity related to two other angles by means of an equation from standard spherical astronomy applied to triple star systems (cosine rules from spherical trigonometry, \citealt{1981ApJ...246..879F}):
%\begin{equation}
%\cos\phi= \cos i_{AB} \cos i_{Bab}+ \sin i_{AB} \sin i_{Bab} \cos(\Omega_{AB}-\Omega_{Bab})
%\end{equation}
%\begin{equation}
%i_{\rm in} - i_{\rm out} \leq \phi \leq i_{\rm in} + i_{\rm out}
%\end{equation}
%Dynamical stability \citep{2001MNRAS.321..398M}.\\
%Kozai invariant: $\Theta = (1-e_{\rm in}^2) \cos^2 \phi$.\\
%$e_{\rm in}$: eccentricity of the inner sub-system Bab (inner eccentricity).\\
%Kozai cycles with tidal friction (KCTF; \citealt{1979A&A....77..145M}) $\Rightarrow$ shrink and circularise the inner orbit.\\
%Kozai-Lidov mechanism cannot be ruled out for HD~188753.\\
%More advanced three-body simulations \citep{2016ComAC...3....6T}.

%%%%%%%%%%%%%%%%%%%%%%%%%%Impact of the input physics%%%%%%%%%%%%%%%%%%%%%%%%%%%

\subsection{Impact of the input physics}

In this section, we investigate the impact of the input physics on the mass and age of star~A. For this, we used the 50 grid models determined in \refsec{sec:mod_opt} as a starting point for the optimisation. Again, we employed a Levenberg-Marquardt minimisation method to obtain the best-fit model for the different sets of input physics considered.

\begin{figure}[tbp]
\centering
% trim=g b d h
\includegraphics[trim = 1.4cm 6.5cm 2.2cm 7.5cm, clip,width=0.5\textwidth]{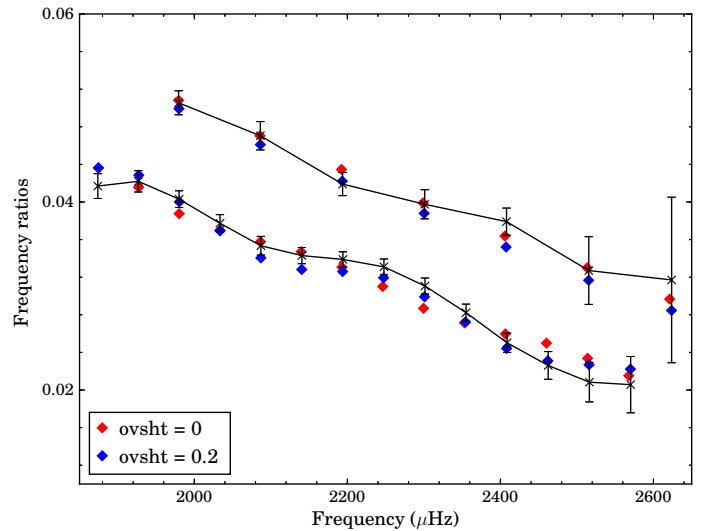}
% for a referee version
%\includegraphics[trim = 1.4cm 6.5cm 2.2cm 7.5cm, clip,width=0.8\textwidth]{6469154_r010_r02_ovsht.ps}
\caption{Frequency separation ratios as a function of frequency for star~A. Red diamonds correspond to our reference model computed without overshooting while blue diamonds correspond to our optimal model computed with overshooting ($\alpha_{\rm ov}$ $=0.2$). The observed ratios are shown as connected points with errors.}
\label{fig:ratios_ovsht}
\end{figure}

We remind the reader that our reference model, labelled `Ref' in \reftab{tab:phys_models}, was computed without overshooting, using the solar mixture of \citeauthor{2009ARA&A..47..481A} (\citeyear{2009ARA&A..47..481A}, hereafter AGSS09) and including microscopic diffusion. In order to check the consistency of the derived stellar parameters from our reference model, we then adopted the following input physics during the minimisation process:
\begin{itemize}
\item Model `Ovsht' computed assuming an overshoot parameter $\alpha_{\rm ov}$ $=0.2$.
\item Model `GS98' computed using the solar mixture of \cite{1998SSRv...85..161G}.
\item Model `Nodiff' computed without microscopic diffusion.
\end{itemize}
For each case, an optimal model was found by fitting the frequency ratios, as illustrated in \reffig{fig:ratios_ovsht}. The corresponding stellar parameters are listed in \reftab{tab:phys_models}.

\begin{table*}[tbp]
\caption{Derived fundamental parameters of star~A for all the models using different input physics.}
\centering
\begin{tabular*}{1.0\textwidth}{@{\extracolsep{\fill}}lccccccccccc} 
\hline
\hline
Model &  $M$ & Age & $Y_0$ & $(Z/X)_0$ & $\alpha_{\rm MLT}$ & $L$ & $R$ & $T_{\rm eff}$ & $\log g$ & ${\rm [Fe/H]}$ & $\chi^2_N$ \\
& ($M_\odot$) &  (Gyr) & & & &  ($L_\odot$) & ($R_\odot$) & (K) & (cgs) & (dex) & \\
\hline
Ref & $0.99 \pm 0.01$ & $10.7 \pm 0.2$ & $0.278 \pm 0.002$ & $0.033 \pm 0.001$ & $1.83 \pm 0.06$ & 1.23 & 1.18 & 5598 & 4.29 & 0.17 & 2.35 \\
Ovsht & $1.00 \pm 0.01$ & $10.9 \pm 0.2$ & $0.277 \pm 0.002$ & $0.034 \pm 0.001$ & $1.87 \pm 0.05$ & 1.22 & 1.18 & 5592 & 4.29 & 0.19 & 1.10 \\ 
GS98 & $0.95 \pm 0.01$ & $11.6 \pm 0.2$ & $0.290 \pm 0.002$ & $0.039 \pm 0.002$ & $1.63 \pm 0.07$ & 1.10 & 1.16 & 5487 & 4.28 & 0.13 & 1.76 \\ 
Nodiff & $1.08 \pm 0.01$ & $10.1 \pm 0.2$ & $0.257 \pm 0.002$ & $0.040 \pm 0.001$ & $1.89 \pm 0.07$ & 1.33 & 1.21 & 5646 & 4.31 & 0.34 & 3.16\vspace*{0.05cm} \\ 
\hline
\end{tabular*}
\tablefoot{\\
Model `Ref' computed without overshooting, using the AGSS09 solar mixture and including microscopic diffusion (see \refsec{sec:input_phys}).\\
Model `Ovsht' computed assuming an overshoot parameter $\alpha_{\rm ov}$ $=0.2$.\\
Model `GS98' computed using the solar mixture of \cite{1998SSRv...85..161G}.\\
Model `Nodiff' computed without microscopic diffusion.
}
\label{tab:phys_models}
\end{table*}

Based on the modelling results, the mass of star~A appears to be in the range $[0.95,1.08]\,M_\odot$. Unfortunately, the large uncertainty on the orbital value of the mass, $M_A =$ $1.05^{+0.10}_{-0.09}\,M_\odot$, does not allow us to discard either of the models. In addition, we find the age of star~A to be in the range 10.1--11.6$\,{\rm Gyr}$. As a result, the estimate of the age also suffers from a large uncertainty due to the physics used in the models. On the other hand, the inclusion of overshooting produces a better fit to the frequency ratios (see \reffig{fig:ratios_ovsht}). For this model, the derived values of the stellar parameters are in good agreement with those from our reference model, thereby reinforcing the results of this study.

We point out that further radial-velocity observations of HD~188753 are expected to provide much more stringent constraints on the stellar masses, which should help to disentangle the different input physics. In particular, the choice of the solar mixture requires a dedicated study, which is beyond the scope of this paper.

%%%%%%%%%%%%%%%%%%%%%%%%%%Dynamical stability%%%%%%%%%%%%%%%%%%%%%%%%%%%

%\subsection{Dynamical stability}

%\citep{2005ApJ...633L.141P,2005ApJ...635L..89P,2007ApJ...654..641J}.\\
%Stability criterion \citep{2001MNRAS.321..398M} $\Rightarrow$ ratio of the pericentre distance of the outer orbit to the semi-major axis of the inner orbit:
%\begin{equation}
%\frac{R_{\rm p}^{\rm out}}{a_{\rm in}} > 2.8 \left(1-0.3\frac{\phi}{\pi}\right) \left((1+q_{\rm out}) \frac{1+e_{\rm out}}{(1-e_{\rm out})^{1/2}}\right)^{2/5}
%\end{equation}
%Outer orbit refers to the AB system while the inner orbit refers to the Bab sub-system.\\
%$R_{\rm p}^{\rm out}=a_{\rm out}(1-e_{\rm out})$ $\Rightarrow$ $a_{\rm out}$, $e_{\rm out}$: outer eccentricity and semi-major axis. \\
%$q_{\rm out}=M_A/(M_{Ba}+M_{Bb})$: outer mass ratio.\\
%$(1-0.3\phi/\pi)$: empirical factor \citep{1972CeMec...6..322H,1999ASIC..522..385M}.\\
%$\phi \in [0,\pi]$ in radians.\\
%Dynamical stability of HD~188753 without having to do more advanced three-body simulations.

%%%%%%%%%%%%%%%%%%%%%%%%%%%%%%%%%%%%%%%%%%%%%%%%%%%%%%%%%%%%%
%%%%%%%%%%%%%%%%%%%%%%%%%%Conclusions%%%%%%%%%%%%%%%%%%%%%%%%%%%
%%%%%%%%%%%%%%%%%%%%%%%%%%%%%%%%%%%%%%%%%%%%%%%%%%%%%%%%%%%%%

\section{Conclusions}\label{sec:conclusions}

Using \kep{} photometry, we report, for the first time, the detection of solar-like oscillations in the two brightest components of a close triple star system, HD~188753. We performed the seismic analysis of the two stars, providing accurate mode frequencies and reliable proxies of the stellar masses and ages. From the modelling, we also derived precise but model-dependent stellar parameters for the two oscillating components. In particular, we found that the mean common age of the system is $10.8 \pm 0.2\,{\rm Gyr}$ while the masses of the two seismic stars are $M_A =$ $0.99 \pm 0.01\,M_\odot$ and $M_{Ba} =$ $0.86 \pm 0.01\,M_\odot$. 
Furthermore, we explored the impact on the derived stellar parameters of varying the input physics of the models. For star~A, the best-fit model was obtained by assuming an overshoot parameter of 0.2.

%Thus, the estimated values of the stellar masses are $M_A =$ $0.99 \pm 0.01\,M_\odot$ and $M_{Ba} =$ $0.86 \pm 0.01\,M_\odot$. In addition, the two stars have a similar age of $10.8 \pm 0.2\,{\rm Gyr}$, as expected from their common origin.

We also performed the first orbital analysis of HD~188753 that combines both the position and radial-velocity measurements of the system. For the wide pair, we derived the masses of the visual components and the parallax of the system, $\pi =$ $21.9 \pm 0.6\,{\rm mas}$. In addition, we derived a precise mass ratio between both stars of the close pair, $M_{Bb}/M_{Ba}  = 0.767 \pm 0.006$. We then found individual masses that are in good agreement with the modelling results. Finally, from the orbital analysis, we estimated the total mass of the system to be $M_{\rm syst} =$ $2.51^{+0.20}_{-0.18}\,M_\odot$.
%Adopting a Bayesian approach, we performed the first analysis of HD~188753 that combines both the astrometric and radial-velocity measurements of the system.

Combining the asteroseismic and astrometric observations of HD~188753 allowed us to better characterise the main features of this triple star system. For example, using the mass ratio of the close pair, we precisely determined the mass of the faintest star, $M_{Bb} =$ $0.66 \pm 0.01\,M_\odot$, from the seismic mass of its companion. We then estimated the total seismic mass of the system to be $M_{\rm syst} =$ $2.51 \pm 0.02\,M_\odot$, which agrees perfectly with the orbital value. Injecting this latter into the Kepler’s third law leads to the best current estimate of the parallax for HD~188753, namely $\pi = 21.9 \pm 0.2\,{\rm mas}$. From our combined analysis, we also derived stringent limits on the relative inclination of the system. With a lower limit at $10.9^\circ \pm 1.5^\circ$, we concluded that HD~188753 does not have a coplanar configuration.

%Double-lined spectroscopic binary (SB2)  $\Rightarrow$ mass ratio determined with high accuracy.\\
%+ independent distance from Hipparcos (and Gaia).\\
%Triple and higher-order systems represent $\sim$20$\%$ of the total stellar population.\\ 
%Close triple system $\Rightarrow$ study of multiple stellar systems.

In this work, we demonstrated that binaries and multiple stellar systems have the potential to constrain fundamental stellar parameters such as the mass and the age. Further observations using asteroseismology and astrometry also promise to improve our knowledge about the dynamical evolution of triple and higher-order systems. In this context, we stress the necessity to prepare catalogues of binaries and multiples for the future space missions TESS \citep{2015JATIS...1a4003R} and PLATO \citep{2014ExA....38..249R}.
%TESS \citep{2015JATIS...1a4003R,2016ApJ...830..138C}

\begin{acknowledgements}
This paper includes data collected by the \kep{} mission. Funding for the \kep{} mission is provided by the NASA Science Mission directorate.
The authors gratefully acknowledge the \kep{} Science Operations Center (SOC) for reprocessing the short-cadence data that enabled the detection of the secondary seismic component.
This research has made use of NASA's Astrophysics Data System Bibliographic Services, the Washington Double Star Catalog maintained at the U.S. Naval Observatory, the SIMBAD database, operated at CDS, Strasbourg, France and the VizieR catalogue access tool, CDS, Strasbourg, France. The original description of the VizieR service was
 published in A\&{}AS 143, 23.
Finally, we also thank the anonymous referee for comments that helped to improve this paper.
\end{acknowledgements}

\bibliographystyle{aa}
\bibliography{fredm}

\begin{thebibliography}{120}
\expandafter\ifx\csname natexlab\endcsname\relax\def\natexlab#1{#1}\fi

\bibitem[{{Alexander} \& {Ferguson}(1994)}]{1994ApJ...437..879A}
{Alexander}, D.~R. \& {Ferguson}, J.~W. 1994, \apj, 437, 879

\bibitem[{{Anderson} {et~al.}(1990){Anderson}, {Duvall}, \&
  {Jefferies}}]{1990ApJ...364..699A}
{Anderson}, E.~R., {Duvall}, Jr., T.~L., \& {Jefferies}, S.~M. 1990, \apj, 364,
  699

\bibitem[{{Angulo} {et~al.}(1999){Angulo}, {Arnould}, {Rayet}, {Descouvemont},
  {Baye}, {Leclercq-Willain}, {Coc}, {Barhoumi}, {Aguer}, {Rolfs}, {Kunz},
  {Hammer}, {Mayer}, {Paradellis}, \& {Kossionides}}]{ang99}
{Angulo}, C., {Arnould}, M., {Rayet}, M., {et~al.} 1999, Nuclear Physics A,
  656, 3

\bibitem[{{Appourchaux}(2014)}]{2014aste.book..123A}
{Appourchaux}, T. 2014, {A crash course on data analysis in asteroseismology},
  ed. P.~L. {Pall{\'e}} \& C.~{Esteban}, 123

\bibitem[{{Appourchaux} {et~al.}(2015){Appourchaux}, {Antia}, {Ball},
  {Creevey}, {Lebreton}, {Verma}, {Vorontsov}, {Campante}, {Davies}, {Gaulme},
  {R{\'e}gulo}, {Horch}, {Howell}, {Everett}, {Ciardi}, {Fossati}, {Miglio},
  {Montalb{\'a}n}, {Chaplin}, {Garc{\'{\i}}a}, \&
  {Gizon}}]{2015A&A...582A..25A}
{Appourchaux}, T., {Antia}, H.~M., {Ball}, W., {et~al.} 2015, \aap, 582, A25

\bibitem[{{Appourchaux} {et~al.}(2014){Appourchaux}, {Antia}, {Benomar},
  {Campante}, {Davies}, {Handberg}, {Howe}, {R{\'e}gulo}, {Belkacem}, {Houdek},
  {Garc{\'{\i}}a}, \& {Chaplin}}]{2014A&A...566A..20A}
{Appourchaux}, T., {Antia}, H.~M., {Benomar}, O., {et~al.} 2014, \aap, 566, A20

\bibitem[{{Appourchaux} {et~al.}(2012{\natexlab{a}}){Appourchaux}, {Benomar},
  {Gruberbauer}, {Chaplin}, {Garc{\'{\i}}a}, {Handberg}, {Verner}, {Antia},
  {Campante}, {Davies}, {Deheuvels}, {Hekker}, {Howe}, {Salabert}, {Bedding},
  {White}, {Houdek}, {Silva Aguirre}, {Elsworth}, {van Cleve}, {Clarke},
  {Hall}, \& {Kjeldsen}}]{2012A&A...537A.134A}
{Appourchaux}, T., {Benomar}, O., {Gruberbauer}, M., {et~al.}
  2012{\natexlab{a}}, \aap, 537, A134

\bibitem[{{Appourchaux} {et~al.}(2012{\natexlab{b}}){Appourchaux}, {Chaplin},
  {Garc{\'{\i}}a}, {Gruberbauer}, {Verner}, {Antia}, {Benomar}, {Campante},
  {Davies}, {Deheuvels}, {Handberg}, {Hekker}, {Howe}, {R{\'e}gulo},
  {Salabert}, {Bedding}, {White}, {Ballot}, {Mathur}, {Silva Aguirre},
  {Elsworth}, {Basu}, {Gilliland}, {Christensen-Dalsgaard}, {Kjeldsen},
  {Uddin}, {Stumpe}, \& {Barclay}}]{2012A&A...543A..54A}
{Appourchaux}, T., {Chaplin}, W.~J., {Garc{\'{\i}}a}, R.~A., {et~al.}
  2012{\natexlab{b}}, \aap, 543, A54

\bibitem[{{Armstrong} {et~al.}(1992){Armstrong}, {Mozurkewich}, {Vivekanand},
  {Simon}, {Denison}, {Johnston}, {Pan}, {Shao}, \&
  {Colavita}}]{1992AJ....104..241A}
{Armstrong}, J.~T., {Mozurkewich}, D., {Vivekanand}, M., {et~al.} 1992, \aj,
  104, 241

\bibitem[{{Asplund} {et~al.}(2009){Asplund}, {Grevesse}, {Sauval}, \&
  {Scott}}]{2009ARA&A..47..481A}
{Asplund}, M., {Grevesse}, N., {Sauval}, A.~J., \& {Scott}, P. 2009, \araa, 47,
  481

\bibitem[{{Baglin} {et~al.}(2006){Baglin}, {Auvergne}, {Barge}, {Deleuil},
  {Catala}, {Michel}, {Weiss}, \& {COROT Team}}]{2006ESASP1306...33B}
{Baglin}, A., {Auvergne}, M., {Barge}, P., {et~al.} 2006, in ESA Special
  Publication, Vol. 1306, The CoRoT Mission Pre-Launch Status - Stellar
  Seismology and Planet Finding, ed. M.~{Fridlund}, A.~{Baglin}, J.~{Lochard},
  \& L.~{Conroy}, 33

\bibitem[{{Baize}(1952)}]{1952JO.....35...27B}
{Baize}, P. 1952, Journal des Observateurs, 35, 27

\bibitem[{{Baize}(1954)}]{1954JO.....37...73B}
{Baize}, P. 1954, Journal des Observateurs, 37, 73

\bibitem[{{Baize}(1957)}]{1957JO.....40..165B}
{Baize}, P. 1957, Journal des Observateurs, 40, 165

\bibitem[{{Ball} \& {Gizon}(2014{\natexlab{a}})}]{2014A&A...568A.123B}
{Ball}, W.~H. \& {Gizon}, L. 2014{\natexlab{a}}, \aap, 568, A123

\bibitem[{{Ball} \& {Gizon}(2014{\natexlab{b}})}]{2014A&A...569C...2B}
{Ball}, W.~H. \& {Gizon}, L. 2014{\natexlab{b}}, \aap, 569, C2

\bibitem[{{Baranne} {et~al.}(1996){Baranne}, {Queloz}, {Mayor}, {Adrianzyk},
  {Knispel}, {Kohler}, {Lacroix}, {Meunier}, {Rimbaud}, \&
  {Vin}}]{1996A&AS..119..373B}
{Baranne}, A., {Queloz}, D., {Mayor}, M., {et~al.} 1996, \aaps, 119, 373

\bibitem[{{Batten}(1973)}]{1973bmss.book.....B}
{Batten}, A.~H. 1973, {Binary and multiple systems of stars}

\bibitem[{{B{\"o}hm-Vitense}(1958)}]{1958ZA.....46..108B}
{B{\"o}hm-Vitense}, E. 1958, \zap, 46, 108

\bibitem[{{Bouchy} \& {Carrier}(2002)}]{2002A&A...390..205B}
{Bouchy}, F. \& {Carrier}, F. 2002, \aap, 390, 205

\bibitem[{{Carrier} \& {Bourban}(2003)}]{2003A&A...406L..23C}
{Carrier}, F. \& {Bourban}, G. 2003, \aap, 406, L23

\bibitem[{{Chaplin} {et~al.}(2014){Chaplin}, {Basu}, {Huber}, {Serenelli},
  {Casagrande}, {Silva Aguirre}, {Ball}, {Creevey}, {Gizon}, {Handberg},
  {Karoff}, {Lutz}, {Marques}, {Miglio}, {Stello}, {Suran}, {Pricopi},
  {Metcalfe}, {Monteiro}, {Molenda-{\.Z}akowicz}, {Appourchaux},
  {Christensen-Dalsgaard}, {Elsworth}, {Garc{\'{\i}}a}, {Houdek}, {Kjeldsen},
  {Bonanno}, {Campante}, {Corsaro}, {Gaulme}, {Hekker}, {Mathur}, {Mosser},
  {R{\'e}gulo}, \& {Salabert}}]{2014ApJS..210....1C}
{Chaplin}, W.~J., {Basu}, S., {Huber}, D., {et~al.} 2014, \apjs, 210, 1

\bibitem[{{Chaplin} {et~al.}(2011){Chaplin}, {Kjeldsen},
  {Christensen-Dalsgaard}, {Basu}, {Miglio}, {Appourchaux}, {Bedding},
  {Elsworth}, {Garc{\'{\i}}a}, {Gilliland}, {Girardi}, {Houdek}, {Karoff},
  {Kawaler}, {Metcalfe}, {Molenda-{\.Z}akowicz}, {Monteiro}, {Thompson},
  {Verner}, {Ballot}, {Bonanno}, {Brand{\~a}o}, {Broomhall}, {Bruntt},
  {Campante}, {Corsaro}, {Creevey}, {Do{\u g}an}, {Esch}, {Gai}, {Gaulme},
  {Hale}, {Handberg}, {Hekker}, {Huber}, {Jim{\'e}nez}, {Mathur}, {Mazumdar},
  {Mosser}, {New}, {Pinsonneault}, {Pricopi}, {Quirion}, {R{\'e}gulo},
  {Salabert}, {Serenelli}, {Silva Aguirre}, {Sousa}, {Stello}, {Stevens},
  {Suran}, {Uytterhoeven}, {White}, {Borucki}, {Brown}, {Jenkins}, {Kinemuchi},
  {Van Cleve}, \& {Klaus}}]{2011Sci...332..213C}
{Chaplin}, W.~J., {Kjeldsen}, H., {Christensen-Dalsgaard}, J., {et~al.} 2011,
  Science, 332, 213

\bibitem[{{Christensen-Dalsgaard}(1984)}]{JCD84b}
{Christensen-Dalsgaard}, J. 1984, in Space Research in Stellar Activity and
  Variability, ed. A.Mangeney \& F.Praderie (Observatoire de Meudon), 11

\bibitem[{{Christensen-Dalsgaard}(2008)}]{jcd08adipls}
{Christensen-Dalsgaard}, J. 2008, \apss, 316, 113

\bibitem[{{Couteau}(1967)}]{1967JO.....50...41C}
{Couteau}, P. 1967, Journal des Observateurs, 50, 41

\bibitem[{{Couteau}(1970)}]{1970AAS....3...51C}
{Couteau}, P. 1970, \aaps, 3, 51

\bibitem[{{Couteau} {et~al.}(1993){Couteau}, {Docobo}, \&
  {Ling}}]{1993AAS..100..305C}
{Couteau}, P., {Docobo}, J.~A., \& {Ling}, J. 1993, \aaps, 100, 305

\bibitem[{{Couteau} \& {Ling}(1991)}]{1991AAS...88..497C}
{Couteau}, P. \& {Ling}, J. 1991, \aaps, 88, 497

\bibitem[{{Creevey} {et~al.}(2017){Creevey}, {Metcalfe}, {Schultheis},
  {Salabert}, {Bazot}, {Th{\'e}venin}, {Mathur}, {Xu}, \&
  {Garc{\'{\i}}a}}]{2017A&A...601A..67C}
{Creevey}, O.~L., {Metcalfe}, T.~S., {Schultheis}, M., {et~al.} 2017, \aap,
  601, A67

\bibitem[{{Davies} {et~al.}(2015){Davies}, {Chaplin}, {Farr}, {Garc{\'{\i}}a},
  {Lund}, {Mathis}, {Metcalfe}, {Appourchaux}, {Basu}, {Benomar}, {Campante},
  {Ceillier}, {Elsworth}, {Handberg}, {Salabert}, \&
  {Stello}}]{2015MNRAS.446.2959D}
{Davies}, G.~R., {Chaplin}, W.~J., {Farr}, W.~M., {et~al.} 2015, \mnras, 446,
  2959

\bibitem[{{Davies} {et~al.}(2014){Davies}, {Handberg}, {Miglio}, {Campante},
  {Chaplin}, \& {Elsworth}}]{Davies2014}
{Davies}, G.~R., {Handberg}, R., {Miglio}, A., {et~al.} 2014, \mnras, 445, L94

\bibitem[{{Davies} {et~al.}(2016){Davies}, {Silva Aguirre}, {Bedding},
  {Handberg}, {Lund}, {Chaplin}, {Huber}, {White}, {Benomar}, {Hekker}, {Basu},
  {Campante}, {Christensen-Dalsgaard}, {Elsworth}, {Karoff}, {Kjeldsen},
  {Lundkvist}, {Metcalfe}, \& {Stello}}]{2016MNRAS.456.2183D}
{Davies}, G.~R., {Silva Aguirre}, V., {Bedding}, T.~R., {et~al.} 2016, \mnras,
  456, 2183

\bibitem[{{Eggenberger} {et~al.}(2007){Eggenberger}, {Udry}, {Mazeh}, {Segal},
  \& {Mayor}}]{2007A&A...466.1179E}
{Eggenberger}, A., {Udry}, S., {Mazeh}, T., {Segal}, Y., \& {Mayor}, M. 2007,
  \aap, 466, 1179

\bibitem[{{ESA}(1997)}]{1997ESASP1200.....E}
{ESA}, ed. 1997, ESA Special Publication, Vol. 1200, {The HIPPARCOS and TYCHO
  catalogues. Astrometric and photometric star catalogues derived from the ESA
  HIPPARCOS Space Astrometry Mission}

\bibitem[{{Fekel}(1981)}]{1981ApJ...246..879F}
{Fekel}, Jr., F.~C. 1981, \apj, 246, 879

\bibitem[{{Formicola} {et~al.}(2004){Formicola}, {Imbriani}, {Costantini},
  {Angulo}, {Bemmerer}, {Bonetti}, {Broggini}, {Corvisiero}, {Cruz},
  {Descouvemont}, {F{\"u}l{\"o}p}, {Gervino}, {Guglielmetti}, {Gustavino},
  {Gy{\"u}rky}, {Jesus}, {Junker}, {Lemut}, {Menegazzo}, {Prati}, {Roca},
  {Rolfs}, {Romano}, {Rossi Alvarez}, {Sch{\"u}mann}, {Somorjai}, {Straniero},
  {Strieder}, {Terrasi}, {Trautvetter}, {Vomiero}, \&
  {Zavatarelli}}]{Formicola2004}
{Formicola}, A., {Imbriani}, G., {Costantini}, H., {et~al.} 2004, Physics
  Letters B, 591, 61

\bibitem[{{Gilliland} {et~al.}(2010{\natexlab{a}}){Gilliland}, {Brown},
  {Christensen-Dalsgaard}, {Kjeldsen}, {Aerts}, {Appourchaux}, {Basu},
  {Bedding}, {Chaplin}, {Cunha}, {De Cat}, {De Ridder}, {Guzik}, {Handler},
  {Kawaler}, {Kiss}, {Kolenberg}, {Kurtz}, {Metcalfe}, {Monteiro}, {Szab{\'o}},
  {Arentoft}, {Balona}, {Debosscher}, {Elsworth}, {Quirion}, {Stello},
  {Su{\'a}rez}, {Borucki}, {Jenkins}, {Koch}, {Kondo}, {Latham}, {Rowe}, \&
  {Steffen}}]{2010PASP..122..131G}
{Gilliland}, R.~L., {Brown}, T.~M., {Christensen-Dalsgaard}, J., {et~al.}
  2010{\natexlab{a}}, \pasp, 122, 131

\bibitem[{{Gilliland} {et~al.}(2010{\natexlab{b}}){Gilliland}, {Jenkins},
  {Borucki}, {Bryson}, {Caldwell}, {Clarke}, {Dotson}, {Haas}, {Hall}, {Klaus},
  {Koch}, {McCauliff}, {Quintana}, {Twicken}, \& {van
  Cleve}}]{2010ApJ...713L.160G}
{Gilliland}, R.~L., {Jenkins}, J.~M., {Borucki}, W.~J., {et~al.}
  2010{\natexlab{b}}, \apjl, 713, L160

\bibitem[{{Grevesse} \& {Sauval}(1998)}]{1998SSRv...85..161G}
{Grevesse}, N. \& {Sauval}, A.~J. 1998, \ssr, 85, 161

\bibitem[{{Griffin}(1967)}]{1967ApJ...148..465G}
{Griffin}, R.~F. 1967, \apj, 148, 465

\bibitem[{{Griffin}(1977)}]{1977Obs....97...15G}
{Griffin}, R.~F. 1977, The Observatory, 97, 15

\bibitem[{{Gruberbauer} {et~al.}(2013){Gruberbauer}, {Guenther}, {MacLeod}, \&
  {Kallinger}}]{2013MNRAS.435..242G}
{Gruberbauer}, M., {Guenther}, D.~B., {MacLeod}, K., \& {Kallinger}, T. 2013,
  \mnras, 435, 242

\bibitem[{{Hartkopf} \& {Mason}(2009)}]{2009AJ....138..813H}
{Hartkopf}, W.~I. \& {Mason}, B.~D. 2009, \aj, 138, 813

\bibitem[{{Hartkopf} {et~al.}(2000){Hartkopf}, {Mason}, {McAlister}, {Roberts},
  {Turner}, {ten Brummelaar}, {Prieto}, {Ling}, \&
  {Franz}}]{2000AJ....119.3084H}
{Hartkopf}, W.~I., {Mason}, B.~D., {McAlister}, H.~A., {et~al.} 2000, \aj, 119,
  3084

\bibitem[{{Hartkopf} {et~al.}(2001{\natexlab{a}}){Hartkopf}, {Mason}, \&
  {Worley}}]{2001AJ....122.3472H}
{Hartkopf}, W.~I., {Mason}, B.~D., \& {Worley}, C.~E. 2001{\natexlab{a}}, \aj,
  122, 3472

\bibitem[{{Hartkopf} {et~al.}(2001{\natexlab{b}}){Hartkopf}, {McAlister}, \&
  {Mason}}]{2001AJ....122.3480H}
{Hartkopf}, W.~I., {McAlister}, H.~A., \& {Mason}, B.~D. 2001{\natexlab{b}},
  \aj, 122, 3480

\bibitem[{{Hartkopf} {et~al.}(1997){Hartkopf}, {McAlister}, {Mason}, {ten
  Brummelaar}, {Roberts}, {Turner}, \& {Wilson}}]{1997AJ....114.1639H}
{Hartkopf}, W.~I., {McAlister}, H.~A., {Mason}, B.~D., {et~al.} 1997, \aj, 114,
  1639

\bibitem[{Hastings(1970)}]{hastings70}
Hastings, W.~K. 1970, Biometrika, 57, 97

\bibitem[{{Heintz}(1975)}]{1975ApJS...29..315H}
{Heintz}, W.~D. 1975, \apjs, 29, 315

\bibitem[{{Heintz}(1978)}]{1978GAM....15.....H}
{Heintz}, W.~D. 1978, Geophysics and Astrophysics Monographs, 15

\bibitem[{{Heintz}(1980)}]{1980ApJS...44..111H}
{Heintz}, W.~D. 1980, \apjs, 44, 111

\bibitem[{{Heintz}(1990)}]{1990ApJS...74..275H}
{Heintz}, W.~D. 1990, \apjs, 74, 275

\bibitem[{{Heintz}(1998)}]{1998ApJS..117..587H}
{Heintz}, W.~D. 1998, \apjs, 117, 587

\bibitem[{{Holden}(1975)}]{1975PASP...87..253H}
{Holden}, F. 1975, \pasp, 87, 253

\bibitem[{{Holden}(1976)}]{1976PASP...88..325H}
{Holden}, F. 1976, \pasp, 88, 325

\bibitem[{{Horch} {et~al.}(1999){Horch}, {Ninkov}, {van Altena}, {Meyer},
  {Girard}, \& {Timothy}}]{1999AJ....117..548H}
{Horch}, E., {Ninkov}, Z., {van Altena}, W.~F., {et~al.} 1999, \aj, 117, 548

\bibitem[{{Horch} {et~al.}(2002){Horch}, {Robinson}, {Meyer}, {van Altena},
  {Ninkov}, \& {Piterman}}]{2002AJ....123.3442H}
{Horch}, E.~P., {Robinson}, S.~E., {Meyer}, R.~D., {et~al.} 2002, \aj, 123,
  3442

\bibitem[{{Hough}(1899)}]{1899AN....149...65H}
{Hough}, G.~W. 1899, Astronomische Nachrichten, 149, 65

\bibitem[{{Huber} {et~al.}(2012){Huber}, {Ireland}, {Bedding}, {Brand{\~a}o},
  {Piau}, {Maestro}, {White}, {Bruntt}, {Casagrande}, {Molenda-{\.Z}akowicz},
  {Silva Aguirre}, {Sousa}, {Barclay}, {Burke}, {Chaplin},
  {Christensen-Dalsgaard}, {Cunha}, {De Ridder}, {Farrington}, {Frasca},
  {Garc{\'{\i}}a}, {Gilliland}, {Goldfinger}, {Hekker}, {Kawaler}, {Kjeldsen},
  {McAlister}, {Metcalfe}, {Miglio}, {Monteiro}, {Pinsonneault}, {Schaefer},
  {Stello}, {Stumpe}, {Sturmann}, {Sturmann}, {ten Brummelaar}, {Thompson},
  {Turner}, \& {Uytterhoeven}}]{2012ApJ...760...32H}
{Huber}, D., {Ireland}, M.~J., {Bedding}, T.~R., {et~al.} 2012, \apj, 760, 32

\bibitem[{{Huber} {et~al.}(2014){Huber}, {Silva Aguirre}, {Matthews},
  {Pinsonneault}, {Gaidos}, {Garc{\'{\i}}a}, {Hekker}, {Mathur}, {Mosser},
  {Torres}, {Bastien}, {Basu}, {Bedding}, {Chaplin}, {Demory}, {Fleming},
  {Guo}, {Mann}, {Rowe}, {Serenelli}, {Smith}, \&
  {Stello}}]{2014ApJS..211....2H}
{Huber}, D., {Silva Aguirre}, V., {Matthews}, J.~M., {et~al.} 2014, \apjs, 211,
  2

\bibitem[{{Iglesias} \& {Rogers}(1996)}]{1996ApJ...464..943I}
{Iglesias}, C.~A. \& {Rogers}, F.~J. 1996, \apj, 464, 943

\bibitem[{{Jenkins} {et~al.}(2010){Jenkins}, {Caldwell}, {Chandrasekaran},
  {Twicken}, {Bryson}, {Quintana}, {Clarke}, {Li}, {Allen}, {Tenenbaum}, {Wu},
  {Klaus}, {Middour}, {Cote}, {McCauliff}, {Girouard}, {Gunter}, {Wohler},
  {Sommers}, {Hall}, {Uddin}, {Wu}, {Bhavsar}, {Van Cleve}, {Pletcher},
  {Dotson}, {Haas}, {Gilliland}, {Koch}, \& {Borucki}}]{2010ApJ...713L..87J}
{Jenkins}, J.~M., {Caldwell}, D.~A., {Chandrasekaran}, H., {et~al.} 2010,
  \apjl, 713, L87

\bibitem[{{Kjeldsen} {et~al.}(2008){Kjeldsen}, {Bedding}, \&
  {Christensen-Dalsgaard}}]{2008ApJ...683L.175K}
{Kjeldsen}, H., {Bedding}, T.~R., \& {Christensen-Dalsgaard}, J. 2008, \apjl,
  683, L175

\bibitem[{{Kjeldsen} {et~al.}(2010){Kjeldsen}, {Christensen-Dalsgaard},
  {Handberg}, {Brown}, {Gilliland}, {Borucki}, \& {Koch}}]{2010AN....331..966K}
{Kjeldsen}, H., {Christensen-Dalsgaard}, J., {Handberg}, R., {et~al.} 2010,
  Astronomische Nachrichten, 331, 966

\bibitem[{{Konacki}(2005)}]{2005Natur.436..230K}
{Konacki}, M. 2005, \nat, 436, 230

\bibitem[{{Kozai}(1962)}]{1962AJ.....67..591K}
{Kozai}, Y. 1962, \aj, 67, 591

\bibitem[{{Labeyrie} {et~al.}(1974){Labeyrie}, {Bonneau}, {Stachnik}, \&
  {Gezari}}]{1974ApJ...194L.147L}
{Labeyrie}, A., {Bonneau}, D., {Stachnik}, R.~V., \& {Gezari}, D.~Y. 1974,
  \apjl, 194, L147

\bibitem[{{Lebreton} \& {Montalb{\'a}n}(2009)}]{Lebreton2009}
{Lebreton}, Y. \& {Montalb{\'a}n}, J. 2009, in IAU Symposium, Vol. 258, IAU
  Symposium, ed. E.~E. {Mamajek}, D.~R. {Soderblom}, \& R.~F.~G. {Wyse},
  419--430

\bibitem[{{Lidov}(1962)}]{1962P&SS....9..719L}
{Lidov}, M.~L. 1962, \planss, 9, 719

\bibitem[{{Lindegren} {et~al.}(2016){Lindegren}, {Lammers}, {Bastian},
  {Hern{\'a}ndez}, {Klioner}, {Hobbs}, {Bombrun}, {Michalik}, {Ramos-Lerate},
  {Butkevich}, {Comoretto}, {Joliet}, {Holl}, {Hutton}, {Parsons},
  {Steidelm{\"u}ller}, {Abbas}, {Altmann}, {Andrei}, {Anton}, {Bach},
  {Barache}, {Becciani}, {Berthier}, {Bianchi}, {Biermann}, {Bouquillon},
  {Bourda}, {Br{\"u}semeister}, {Bucciarelli}, {Busonero}, {Carlucci},
  {Casta{\~n}eda}, {Charlot}, {Clotet}, {Crosta}, {Davidson}, {de Felice},
  {Drimmel}, {Fabricius}, {Fienga}, {Figueras}, {Fraile}, {Gai}, {Garralda},
  {Geyer}, {Gonz{\'a}lez-Vidal}, {Guerra}, {Hambly}, {Hauser}, {Jordan},
  {Lattanzi}, {Lenhardt}, {Liao}, {L{\"o}ffler}, {McMillan}, {Mignard}, {Mora},
  {Morbidelli}, {Portell}, {Riva}, {Sarasso}, {Serraller}, {Siddiqui}, {Smart},
  {Spagna}, {Stampa}, {Steele}, {Taris}, {Torra}, {van Reeven}, {Vecchiato},
  {Zschocke}, {de Bruijne}, {Gracia}, {Raison}, {Lister}, {Marchant},
  {Messineo}, {Soffel}, {Osorio}, {de Torres}, \&
  {O'Mullane}}]{2016A&A...595A...4L}
{Lindegren}, L., {Lammers}, U., {Bastian}, U., {et~al.} 2016, \aap, 595, A4

\bibitem[{{Lomb}(1976)}]{1976Ap&SS..39..447L}
{Lomb}, N.~R. 1976, \apss, 39, 447

\bibitem[{{Lund} {et~al.}(2017){Lund}, {Silva Aguirre}, {Davies}, {Chaplin},
  {Christensen-Dalsgaard}, {Houdek}, {White}, {Bedding}, {Ball}, {Huber},
  {Antia}, {Lebreton}, {Latham}, {Handberg}, {Verma}, {Basu}, {Casagrande},
  {Justesen}, {Kjeldsen}, \& {Mosumgaard}}]{2017ApJ...835..172L}
{Lund}, M.~N., {Silva Aguirre}, V., {Davies}, G.~R., {et~al.} 2017, \apj, 835,
  172

\bibitem[{{Marques} {et~al.}(2013){Marques}, {Goupil}, {Lebreton}, {Talon},
  {Palacios}, {Belkacem}, {Ouazzani}, {Mosser}, {Moya}, {Morel}, {Pichon},
  {Mathis}, {Zahn}, {Turck-Chi{\`e}ze}, \& {Nghiem}}]{2013A&A...549A..74M}
{Marques}, J.~P., {Goupil}, M.~J., {Lebreton}, Y., {et~al.} 2013, \aap, 549,
  A74

\bibitem[{{Mason} {et~al.}(2011){Mason}, {Hartkopf}, {Raghavan}, {Subasavage},
  {Roberts}, {Turner}, \& {ten Brummelaar}}]{2011AJ....142..176M}
{Mason}, B.~D., {Hartkopf}, W.~I., {Raghavan}, D., {et~al.} 2011, \aj, 142, 176

\bibitem[{{Mason} {et~al.}(2004){Mason}, {Hartkopf}, {Wycoff}, {Rafferty},
  {Urban}, \& {Flagg}}]{2004AJ....128.3012M}
{Mason}, B.~D., {Hartkopf}, W.~I., {Wycoff}, G.~L., {et~al.} 2004, \aj, 128,
  3012

\bibitem[{{Mazeh} {et~al.}(2009){Mazeh}, {Tsodikovich}, {Segal}, {Zucker},
  {Eggenberger}, {Udry}, \& {Mayor}}]{2009MNRAS.399..906M}
{Mazeh}, T., {Tsodikovich}, Y., {Segal}, Y., {et~al.} 2009, \mnras, 399, 906

\bibitem[{{Mazumdar} {et~al.}(2014){Mazumdar}, {Monteiro}, {Ballot}, {Antia},
  {Basu}, {Houdek}, {Mathur}, {Cunha}, {Silva Aguirre}, {Garc{\'{\i}}a},
  {Salabert}, {Verner}, {Christensen-Dalsgaard}, {Metcalfe}, {Sanderfer},
  {Seader}, {Smith}, \& {Chaplin}}]{Mazumdar2014}
{Mazumdar}, A., {Monteiro}, M.~J.~P.~F.~G., {Ballot}, J., {et~al.} 2014, \apj,
  782, 18

\bibitem[{{Metcalfe} {et~al.}(2012){Metcalfe}, {Chaplin}, {Appourchaux},
  {Garc{\'{\i}}a}, {Basu}, {Brand{\~a}o}, {Creevey}, {Deheuvels}, {Do{\v g}an},
  {Eggenberger}, {Karoff}, {Miglio}, {Stello}, {Y{\i}ld{\i}z}, {{\c C}elik},
  {Antia}, {Benomar}, {Howe}, {R{\'e}gulo}, {Salabert}, {Stahn}, {Bedding},
  {Davies}, {Elsworth}, {Gizon}, {Hekker}, {Mathur}, {Mosser}, {Bryson},
  {Still}, {Christensen-Dalsgaard}, {Gilliland}, {Kawaler}, {Kjeldsen},
  {Ibrahim}, {Klaus}, \& {Li}}]{2012ApJ...748L..10M}
{Metcalfe}, T.~S., {Chaplin}, W.~J., {Appourchaux}, T., {et~al.} 2012, \apjl,
  748, L10

\bibitem[{{Metcalfe} {et~al.}(2015){Metcalfe}, {Creevey}, \&
  {Davies}}]{2015ApJ...811L..37M}
{Metcalfe}, T.~S., {Creevey}, O.~L., \& {Davies}, G.~R. 2015, \apjl, 811, L37

\bibitem[{{Metcalfe} {et~al.}(2014){Metcalfe}, {Creevey}, {Do{\u g}an},
  {Mathur}, {Xu}, {Bedding}, {Chaplin}, {Christensen-Dalsgaard}, {Karoff},
  {Trampedach}, {Benomar}, {Brown}, {Buzasi}, {Campante}, {{\c C}elik},
  {Cunha}, {Davies}, {Deheuvels}, {Derekas}, {Di Mauro}, {Garc{\'{\i}}a},
  {Guzik}, {Howe}, {MacGregor}, {Mazumdar}, {Montalb{\'a}n}, {Monteiro},
  {Salabert}, {Serenelli}, {Stello}, {St\c{e}{\'s}licki}, {Suran},
  {Y{\i}ld{\i}z}, {Aksoy}, {Elsworth}, {Gruberbauer}, {Guenther}, {Lebreton},
  {Molaverdikhani}, {Pricopi}, {Simoniello}, \& {White}}]{2014ApJS..214...27M}
{Metcalfe}, T.~S., {Creevey}, O.~L., {Do{\u g}an}, G., {et~al.} 2014, \apjs,
  214, 27

\bibitem[{{Metropolis} {et~al.}(1953){Metropolis}, {Rosenbluth}, {Rosenbluth},
  {Teller}, \& {Teller}}]{1953JChPh..21.1087M}
{Metropolis}, N., {Rosenbluth}, A.~W., {Rosenbluth}, M.~N., {Teller}, A.~H., \&
  {Teller}, E. 1953, \jcp, 21, 1087

\bibitem[{{Michaud} \& {Proffitt}(1993)}]{1993ASPC...40..246M}
{Michaud}, G. \& {Proffitt}, C.~R. 1993, in Astronomical Society of the Pacific
  Conference Series, Vol.~40, IAU Colloq. 137: Inside the Stars, ed. W.~W.
  {Weiss} \& A.~{Baglin}, 246--259

\bibitem[{{Miglio} {et~al.}(2014){Miglio}, {Chaplin}, {Farmer}, {Kolb},
  {Girardi}, {Elsworth}, {Appourchaux}, \& {Handberg}}]{2014ApJ...784L...3M}
{Miglio}, A., {Chaplin}, W.~J., {Farmer}, R., {et~al.} 2014, \apjl, 784, L3

\bibitem[{{Miglio} {et~al.}(2013){Miglio}, {Chiappini}, {Morel}, {Barbieri},
  {Chaplin}, {Girardi}, {Montalb{\'a}n}, {Valentini}, {Mosser}, {Baudin},
  {Casagrande}, {Fossati}, {Silva Aguirre}, \& {Baglin}}]{2013MNRAS.429..423M}
{Miglio}, A., {Chiappini}, C., {Morel}, T., {et~al.} 2013, \mnras, 429, 423

\bibitem[{{Morel} \& {Lebreton}(2008)}]{2008Ap&SS.316...61M}
{Morel}, P. \& {Lebreton}, Y. 2008, \apss, 316, 61

\bibitem[{{Mosser} {et~al.}(2013){Mosser}, {Michel}, {Belkacem}, {Goupil},
  {Baglin}, {Barban}, {Provost}, {Samadi}, {Auvergne}, \&
  {Catala}}]{Mosser2013}
{Mosser}, B., {Michel}, E., {Belkacem}, K., {et~al.} 2013, \aap, 550, A126

\bibitem[{{Muller}(1955)}]{1955JO.....38..221M}
{Muller}, P. 1955, Journal des Observateurs, 38, 221

\bibitem[{{Ot{\'{\i}} Floranes} {et~al.}(2005){Ot{\'{\i}} Floranes},
  {Christensen-Dalsgaard}, \& {Thompson}}]{Floranes2005}
{Ot{\'{\i}} Floranes}, H., {Christensen-Dalsgaard}, J., \& {Thompson}, M.~J.
  2005, \mnras, 356, 671

\bibitem[{{Planck Collaboration} {et~al.}(2014){Planck Collaboration}, {Ade},
  {Aghanim}, {Armitage-Caplan}, {Arnaud}, {Ashdown}, {Atrio-Barandela},
  {Aumont}, {Baccigalupi}, {Banday}, \& et~al.}]{2014A&A...571A..16P}
{Planck Collaboration}, {Ade}, P.~A.~R., {Aghanim}, N., {et~al.} 2014, \aap,
  571, A16

\bibitem[{{Pourbaix}(2008)}]{2008IAUS..248...59P}
{Pourbaix}, D. 2008, in IAU Symposium, Vol. 248, A Giant Step: from Milli- to
  Micro-arcsecond Astrometry, ed. W.~J. {Jin}, I.~{Platais}, \& M.~A.~C.
  {Perryman}, 59--65

\bibitem[{{Pourbaix} \& {Boffin}(2016)}]{2016A&A...586A..90P}
{Pourbaix}, D. \& {Boffin}, H.~M.~J. 2016, \aap, 586, A90

\bibitem[{{Rauer} {et~al.}(2014){Rauer}, {Catala}, {Aerts}, {Appourchaux},
  {Benz}, {Brandeker}, {Christensen-Dalsgaard}, {Deleuil}, {Gizon}, {Goupil},
  {G{\"u}del}, {Janot-Pacheco}, {Mas-Hesse}, {Pagano}, {Piotto}, {Pollacco},
  {Santos}, {Smith}, {Su{\'a}rez}, {Szab{\'o}}, {Udry}, {Adibekyan}, {Alibert},
  {Almenara}, {Amaro-Seoane}, {Eiff}, {Asplund}, {Antonello}, {Barnes},
  {Baudin}, {Belkacem}, {Bergemann}, {Bihain}, {Birch}, {Bonfils}, {Boisse},
  {Bonomo}, {Borsa}, {Brand{\~a}o}, {Brocato}, {Brun}, {Burleigh}, {Burston},
  {Cabrera}, {Cassisi}, {Chaplin}, {Charpinet}, {Chiappini}, {Church},
  {Csizmadia}, {Cunha}, {Damasso}, {Davies}, {Deeg}, {D{\'{\i}}az}, {Dreizler},
  {Dreyer}, {Eggenberger}, {Ehrenreich}, {Eigm{\"u}ller}, {Erikson}, {Farmer},
  {Feltzing}, {de Oliveira Fialho}, {Figueira}, {Forveille}, {Fridlund},
  {Garc{\'{\i}}a}, {Giommi}, {Giuffrida}, {Godolt}, {Gomes da Silva},
  {Granzer}, {Grenfell}, {Grotsch-Noels}, {G{\"u}nther}, {Haswell}, {Hatzes},
  {H{\'e}brard}, {Hekker}, {Helled}, {Heng}, {Jenkins}, {Johansen},
  {Khodachenko}, {Kislyakova}, {Kley}, {Kolb}, {Krivova}, {Kupka}, {Lammer},
  {Lanza}, {Lebreton}, {Magrin}, {Marcos-Arenal}, {Marrese}, {Marques},
  {Martins}, {Mathis}, {Mathur}, {Messina}, {Miglio}, {Montalban}, {Montalto},
  {Monteiro}, {Moradi}, {Moravveji}, {Mordasini}, {Morel}, {Mortier},
  {Nascimbeni}, {Nelson}, {Nielsen}, {Noack}, {Norton}, {Ofir}, {Oshagh},
  {Ouazzani}, {P{\'a}pics}, {Parro}, {Petit}, {Plez}, {Poretti}, {Quirrenbach},
  {Ragazzoni}, {Raimondo}, {Rainer}, {Reese}, {Redmer}, {Reffert},
  {Rojas-Ayala}, {Roxburgh}, {Salmon}, {Santerne}, {Schneider}, {Schou},
  {Schuh}, {Schunker}, {Silva-Valio}, {Silvotti}, {Skillen}, {Snellen}, {Sohl},
  {Sousa}, {Sozzetti}, {Stello}, {Strassmeier}, {{\v S}vanda}, {Szab{\'o}},
  {Tkachenko}, {Valencia}, {Van Grootel}, {Vauclair}, {Ventura}, {Wagner},
  {Walton}, {Weingrill}, {Werner}, {Wheatley}, \&
  {Zwintz}}]{2014ExA....38..249R}
{Rauer}, H., {Catala}, C., {Aerts}, C., {et~al.} 2014, Experimental Astronomy,
  38, 249

\bibitem[{{Ricker} {et~al.}(2015){Ricker}, {Winn}, {Vanderspek}, {Latham},
  {Bakos}, {Bean}, {Berta-Thompson}, {Brown}, {Buchhave}, {Butler}, {Butler},
  {Chaplin}, {Charbonneau}, {Christensen-Dalsgaard}, {Clampin}, {Deming},
  {Doty}, {De Lee}, {Dressing}, {Dunham}, {Endl}, {Fressin}, {Ge}, {Henning},
  {Holman}, {Howard}, {Ida}, {Jenkins}, {Jernigan}, {Johnson}, {Kaltenegger},
  {Kawai}, {Kjeldsen}, {Laughlin}, {Levine}, {Lin}, {Lissauer}, {MacQueen},
  {Marcy}, {McCullough}, {Morton}, {Narita}, {Paegert}, {Palle}, {Pepe},
  {Pepper}, {Quirrenbach}, {Rinehart}, {Sasselov}, {Sato}, {Seager},
  {Sozzetti}, {Stassun}, {Sullivan}, {Szentgyorgyi}, {Torres}, {Udry}, \&
  {Villasenor}}]{2015JATIS...1a4003R}
{Ricker}, G.~R., {Winn}, J.~N., {Vanderspek}, R., {et~al.} 2015, Journal of
  Astronomical Telescopes, Instruments, and Systems, 1, 014003

\bibitem[{{Rogers} \& {Nayfonov}(2002)}]{2002ApJ...576.1064R}
{Rogers}, F.~J. \& {Nayfonov}, A. 2002, \apj, 576, 1064

\bibitem[{{Roxburgh}(2005)}]{2005A&A...434..665R}
{Roxburgh}, I.~W. 2005, \aap, 434, 665

\bibitem[{{Roxburgh} \& {Vorontsov}(2003)}]{Roxburgh2003a}
{Roxburgh}, I.~W. \& {Vorontsov}, S.~V. 2003, \aap, 411, 215

\bibitem[{{Scargle}(1982)}]{1982ApJ...263..835S}
{Scargle}, J.~D. 1982, \apj, 263, 835

\bibitem[{{Silva Aguirre} {et~al.}(2015){Silva Aguirre}, {Davies}, {Basu},
  {Christensen-Dalsgaard}, {Creevey}, {Metcalfe}, {Bedding}, {Casagrande},
  {Handberg}, {Lund}, {Nissen}, {Chaplin}, {Huber}, {Serenelli}, {Stello}, {Van
  Eylen}, {Campante}, {Elsworth}, {Gilliland}, {Hekker}, {Karoff}, {Kawaler},
  {Kjeldsen}, \& {Lundkvist}}]{2015MNRAS.452.2127S}
{Silva Aguirre}, V., {Davies}, G.~R., {Basu}, S., {et~al.} 2015, \mnras, 452,
  2127

\bibitem[{{S{\"o}derhjelm}(1999)}]{1999AA...341..121S}
{S{\"o}derhjelm}, S. 1999, \aap, 341, 121

\bibitem[{{Sterzik} \& {Tokovinin}(2002)}]{2002A&A...384.1030S}
{Sterzik}, M.~F. \& {Tokovinin}, A.~A. 2002, \aap, 384, 1030

\bibitem[{{Struve}(1837)}]{1837sdmm.book.....S}
{Struve}, F.~G.~W. 1837, {Stellarum duplicium et multiplicium mensurae
  micrometricae per magnum Fraunhoferi tubum annis a 1824 ad 1837 in Specula
  Dorpatensi institutae...}

\bibitem[{{Tassoul}(1980)}]{1980ApJS...43..469T}
{Tassoul}, M. 1980, \apjs, 43, 469

\bibitem[{{Tohline}(2002)}]{Tohline1992}
{Tohline}, J.~E. 2002, \araa, 40, 349

\bibitem[{{Tokovinin}(1980)}]{1980ATsir1097....3T}
{Tokovinin}, A.~A. 1980, Astronomicheskij Tsirkulyar, 1097, 3

\bibitem[{{Tokovinin}(1985)}]{1985AAS...61..483T}
{Tokovinin}, A.~A. 1985, \aaps, 61, 483

\bibitem[{{Toonen} {et~al.}(2016){Toonen}, {Hamers}, \& {Portegies
  Zwart}}]{2016ComAC...3....6T}
{Toonen}, S., {Hamers}, A., \& {Portegies Zwart}, S. 2016, Computational
  Astrophysics and Cosmology, 3, 6

\bibitem[{{Trampedach} {et~al.}(2014){Trampedach}, {Stein},
  {Christensen-Dalsgaard}, {Nordlund}, \& {Asplund}}]{2014MNRAS.445.4366T}
{Trampedach}, R., {Stein}, R.~F., {Christensen-Dalsgaard}, J., {Nordlund},
  {\AA}., \& {Asplund}, M. 2014, \mnras, 445, 4366

\bibitem[{{van Biesbroeck}(1918)}]{1918AJ.....31..169V}
{van Biesbroeck}, G. 1918, \aj, 31, 169

\bibitem[{{van Biesbroeck}(1927)}]{1927PYerO...5....1V}
{van Biesbroeck}, G. 1927, Publications of the Yerkes Observatory, 5, 1.vii

\bibitem[{{van Biesbroeck}(1974)}]{1974ApJS...28..413V}
{van Biesbroeck}, G. 1974, \apjs, 28, 413

\bibitem[{{van den Bos}(1959)}]{1959ApJS....4...45V}
{van den Bos}, W.~H. 1959, \apjs, 4, 45

\bibitem[{{van den Bos}(1963)}]{1963AJ.....68...57V}
{van den Bos}, W.~H. 1963, \aj, 68, 57

\bibitem[{{van Leeuwen}(2007)}]{2007AA...474..653V}
{van Leeuwen}, F. 2007, \aap, 474, 653

\bibitem[{{Verner} {et~al.}(2011){Verner}, {Elsworth}, {Chaplin}, {Campante},
  {Corsaro}, {Gaulme}, {Hekker}, {Huber}, {Karoff}, {Mathur}, {Mosser},
  {Appourchaux}, {Ballot}, {Bedding}, {Bonanno}, {Broomhall}, {Garc{\'{\i}}a},
  {Handberg}, {New}, {Stello}, {R{\'e}gulo}, {Roxburgh}, {Salabert}, {White},
  {Caldwell}, {Christiansen}, \& {Fanelli}}]{2011MNRAS.415.3539V}
{Verner}, G.~A., {Elsworth}, Y., {Chaplin}, W.~J., {et~al.} 2011, \mnras, 415,
  3539

\bibitem[{{Vogt} {et~al.}(1994){Vogt}, {Allen}, {Bigelow}, {Bresee}, {Brown},
  {Cantrall}, {Conrad}, {Couture}, {Delaney}, {Epps}, {Hilyard}, {Hilyard},
  {Horn}, {Jern}, {Kanto}, {Keane}, {Kibrick}, {Lewis}, {Osborne},
  {Pardeilhan}, {Pfister}, {Ricketts}, {Robinson}, {Stover}, {Tucker}, {Ward},
  \& {Wei}}]{1994SPIE.2198..362V}
{Vogt}, S.~S., {Allen}, S.~L., {Bigelow}, B.~C., {et~al.} 1994, in \procspie,
  Vol. 2198, Instrumentation in Astronomy VIII, ed. D.~L. {Crawford} \& E.~R.
  {Craine}, 362

\bibitem[{{White} {et~al.}(2011){White}, {Bedding}, {Stello},
  {Christensen-Dalsgaard}, {Huber}, \& {Kjeldsen}}]{2011ApJ...743..161W}
{White}, T.~R., {Bedding}, T.~R., {Stello}, D., {et~al.} 2011, \apj, 743, 161

\bibitem[{{White} {et~al.}(2017){White}, {Benomar}, {Silva Aguirre}, {Ball},
  {Bedding}, {Chaplin}, {Christensen-Dalsgaard}, {Garcia}, {Gizon}, {Stello},
  {Aigrain}, {Antia}, {Appourchaux}, {Bazot}, {Campante}, {Creevey}, {Davies},
  {Elsworth}, {Gaulme}, {Handberg}, {Hekker}, {Houdek}, {Howe}, {Huber},
  {Karoff}, {Marques}, {Mathur}, {McQuillan}, {Metcalfe}, {Mosser}, {Nielsen},
  {R{\'e}gulo}, {Salabert}, \& {Stahn}}]{2017A&A...601A..82W}
{White}, T.~R., {Benomar}, O., {Silva Aguirre}, V., {et~al.} 2017, \aap, 601,
  A82

\bibitem[{{White} {et~al.}(2013){White}, {Huber}, {Maestro}, {Bedding},
  {Ireland}, {Baron}, {Boyajian}, {Che}, {Monnier}, {Pope}, {Roettenbacher},
  {Stello}, {Tuthill}, {Farrington}, {Goldfinger}, {McAlister}, {Schaefer},
  {Sturmann}, {Sturmann}, {ten Brummelaar}, \& {Turner}}]{2013MNRAS.433.1262W}
{White}, T.~R., {Huber}, D., {Maestro}, V., {et~al.} 2013, \mnras, 433, 1262

\bibitem[{{Worley}(1962)}]{1962AJ.....67..403W}
{Worley}, C.~E. 1962, \aj, 67, 403

\end{thebibliography}

\Online

\appendix

%%%%%%%%%%%%%%%%%%%%%%%%%%%%%%%%%%%%%%%%%%%%%%%%%%%%%%%%%
%%%%%%%%%%%%%%%%%%%%%%%%%%Orbital data%%%%%%%%%%%%%%%%%%%%%%%%%%%
%%%%%%%%%%%%%%%%%%%%%%%%%%%%%%%%%%%%%%%%%%%%%%%%%%%%%%%%%

\onecolumn

\section{Orbital data}\label{app:orb-data}

\LTcapwidth=\textwidth

\begin{longtable}{c c c c}
\caption{Relative positions of the two visual components from micrometric observations. \label{tab:micro_data}}\\
\hline
\hline
Date & Angle & Separation & Reference \\
(Bess. yr.) & (degrees) & (arcsec) & \\
\hline
\endfirsthead
\caption{continued.}\\
\hline
\hline
Date & Angle & Separation & Reference \\
(Bess. yr.) & (degrees) & (arcsec) & \\
\hline
\endhead
\hline
\endfoot
\hline
1895.6750 & 77.40\tablefootmark{a} & 0.340 & \cite{1899AN....149...65H}		\\
1895.7010 & 79.40\tablefootmark{a} & 0.300 & \cite{1899AN....149...65H}		\\
%1895.6700 & 78.40 & 0.360 & {\red [temp]}		\\
%1895.6880 & 78.40 & 0.320 & \cite{1918AJ.....31..169V}		\\
1905.7010 & 149.9 & 0.280 & \cite{1918AJ.....31..169V}		\\
1906.6730 & 158.9 & 0.260 & \cite{1918AJ.....31..169V}		\\
%1906.7100 & 160.6 & 0.230 & {\red [temp]}		\\
1906.7530 & 162.3\tablefootmark{a} & 0.250 & \cite{1918AJ.....31..169V}		\\
1911.5110 & 282.6\tablefootmark{a} & 0.160 & \cite{1918AJ.....31..169V}		\\
1912.6570 & 346.5\tablefootmark{a} & 0.150 & \cite{1918AJ.....31..169V}		\\
%1912.6600 & 346.5 & 0.150 & {\red [temp]}		\\
1914.4900 & 22.80 & 0.200 & \cite{1927PYerO...5....1V}		\\
1917.5200 & 55.60 & 0.310 & \cite{1927PYerO...5....1V}		\\
1917.5680 & 55.10 & 0.300 & \cite{1927PYerO...5....1V}		\\
%1917.5700 & 56.40 & 0.300 & {\red [temp]}		\\
%1917.6050 & 57.20 & 0.280 & \cite{1927PYerO...5....1V}		\\
1917.6090 & 58.30 & 0.290 & \cite{1927PYerO...5....1V}		\\
1917.6200 & 60.50 & 0.270 & \cite{1927PYerO...5....1V}		\\
1917.6230 & 54.80 & 0.270 & \cite{1927PYerO...5....1V}		\\
1918.2820 & 63.20 & 0.300 & \cite{1927PYerO...5....1V}		\\
%1918.4040 & 66.20 & 0.320 & {\red [temp]}		\\
%1918.4050 & 66.20 & 0.320 & \cite{1927PYerO...5....1V}		\\
1918.4190 & 70.20 & 0.310 & \cite{1927PYerO...5....1V}		\\
1918.4680 & 67.10 & 0.320 & \cite{1927PYerO...5....1V}		\\
1918.5010 & 68.30 & 0.320 & \cite{1927PYerO...5....1V}		\\
1918.5400 & 65.60 & 0.340 & \cite{1927PYerO...5....1V}		\\
1918.6400 & 61.00 & 0.290 & \cite{1927PYerO...5....1V}		\\
1919.6300 & 67.00 & 0.260 & \cite{1927PYerO...5....1V}		\\
1920.3300 & 68.60 & 0.410 & \cite{1927PYerO...5....1V}		\\
1920.4120 & 75.20 & 0.360 & \cite{1927PYerO...5....1V}		\\
%1920.5300 & 78.20 & 0.350 & {\red [temp]}		\\
%1920.5340 & 78.30 & 0.350 & \cite{1927PYerO...5....1V}		\\
1920.5660 & 79.20 & 0.320 & \cite{1927PYerO...5....1V}		\\
1920.6230 & 80.40 & 0.380 & \cite{1927PYerO...5....1V}		\\
1921.5600 & 77.20 & 0.340 & \cite{1927PYerO...5....1V}		\\
1921.6500 & 88.00 & 0.310 & \cite{1927PYerO...5....1V}		\\
1923.2400 & 87.00 & 0.370 & \cite{1927PYerO...5....1V}		\\
1923.4980 & 92.20 & 0.400 & \cite{1927PYerO...5....1V}		\\
%1923.5200 & 91.80 & 0.380 & {\red [temp]}		\\
1923.5470 & 89.70 & 0.370 & \cite{1927PYerO...5....1V}		\\
%1923.5690 & 91.60 & 0.380 & \cite{1927PYerO...5....1V}		\\
1923.6620 & 93.00 & 0.370 & \cite{1927PYerO...5....1V}		\\
1924.5030 & 95.40 & 0.320 & \cite{1927PYerO...5....1V}		\\
%1924.5980 & 96.00 & 0.310 & \cite{1927PYerO...5....1V}		\\
%1924.6000 & 96.00 & 0.310 & {\red [temp]}		\\
1924.6420 & 96.60 & 0.290 & \cite{1927PYerO...5....1V}		\\
1924.6480 & 96.10 & 0.310 & \cite{1927PYerO...5....1V}		\\
1925.3780 & 99.60 & 0.340 & \cite{1927PYerO...5....1V}		\\
%1925.4200 & 101.2 & 0.330 & {\red [temp]}		\\
%1925.4200 & 101.2 & 0.360 & {\red [temp]}		\\
%1925.4210 & 101.2 & 0.330 & \cite{1927PYerO...5....1V}		\\
1925.4630 & 102.8 & 0.320 & \cite{1927PYerO...5....1V}		\\
1948.6810 & 84.10 & 0.360 & \cite{1952JO.....35...27B}		\\
1948.7220 & 86.50 & 0.350 & \cite{1952JO.....35...27B}		\\
1948.7280 & 86.30 & 0.370 & \cite{1952JO.....35...27B}		\\
1948.7310 & 85.10 & 0.360 & \cite{1952JO.....35...27B}		\\
1953.7630 & 129.2 & 0.310 & \cite{1954JO.....37...73B}		\\
1953.7750 & 118.5 & 0.290 & \cite{1954JO.....37...73B}		\\
1953.7770 & 119.0 & 0.310 & \cite{1954JO.....37...73B}		\\
1954.7900 & 123.7 & 0.300 & \cite{1955JO.....38..221M}		\\
1954.7900 & 121.1 & 0.280 & \cite{1955JO.....38..221M}		\\
1954.8200 & 123.0 & 0.250 & \cite{1955JO.....38..221M}		\\
1956.7750 & 146.1 & 0.240 & \cite{1957JO.....40..165B}		\\
1956.7770 & 144.5 & 0.280 & \cite{1957JO.....40..165B}		\\
1956.8040 & 146.0 & 0.240 & \cite{1957JO.....40..165B}		\\
1956.8070 & 145.4 & 0.220 & \cite{1957JO.....40..165B}		\\
1957.6090 & 157.6 & 0.270 & \cite{1959ApJS....4...45V}		\\
1957.6200 & 155.4 & 0.280 & \cite{1959ApJS....4...45V}		\\
1957.6280 & 161.6 & 0.260 & \cite{1959ApJS....4...45V}		\\
1960.6880 & 213.6 & 0.180 & \cite{1962AJ.....67..403W}		\\
1960.7050 & 215.7 & 0.180 & \cite{1962AJ.....67..403W}		\\
1960.7570 & 218.9 & 0.180 & \cite{1962AJ.....67..403W}		\\
1960.8030 & 214.0 & 0.160 & \cite{1962AJ.....67..403W}		\\
1962.6720 & 282.5 & 0.130 & \cite{1963AJ.....68...57V}		\\
1966.7190 & 36.60 & 0.300 & \cite{1967JO.....50...41C}		\\
1966.7460 & 33.10 & 0.260 & \cite{1967JO.....50...41C}		\\
1966.7560 & 32.80 & 0.250 & \cite{1967JO.....50...41C}		\\
1969.6510 & 59.20 & 0.350 & \cite{1970AAS....3...51C}		\\
1969.6700 & 64.00 & 0.330 & \cite{1970AAS....3...51C}		\\
1969.6980 & 60.50 & 0.360 & \cite{1970AAS....3...51C}		\\
1972.4700 & 79.10 & 0.340 & \cite{1974ApJS...28..413V}		\\
1972.5410 & 81.40 & 0.320 & \cite{1974ApJS...28..413V}		\\
1974.6000 & 92.50 & 0.350 & \cite{1975ApJS...29..315H}		\\
1974.6400 & 91.40 & 0.350 & \cite{1975PASP...87..253H}		\\
1975.5620 & 95.60 & 0.290 & \cite{1976PASP...88..325H}		\\
1978.6200 & 112.1 & 0.320 & \cite{1980ApJS...44..111H}		\\
1987.5400 & 252.2 & 0.120 & \cite{1990ApJS...74..275H}		\\
1988.5600 & 289.9 & 0.100 & \cite{1990ApJS...74..275H}		\\
1989.4800 & 334.2 & 0.130 & \cite{1990ApJS...74..275H}		\\
1990.5250 & 12.10 & 0.156 & \cite{1991AAS...88..497C}		\\
1990.5290 & 12.80 & 0.189 & \cite{1991AAS...88..497C}		\\
1992.5330 & 41.60 & 0.240 & \cite{1993AAS..100..305C}		\\
1992.5330 & 41.90 & 0.245 & \cite{1993AAS..100..305C}		\\
1992.5530 & 41.60 & 0.237 & \cite{1993AAS..100..305C}		\\
1997.7500 & 79.80 & 0.370 & \cite{1998ApJS..117..587H}		\\
\hline
\end{longtable}
\tablefoot{%
\tablefoottext{a}{These values were corrected for a $180^\circ$ ambiguity in the quadrant determination.}
}

%\twocolumn

%\begin{table*}[!h]
\begin{table}[!h]
\caption{Relative positions of the two visual components from interferometric observations.}
\centering
\begin{tabular}{c c c c}
\hline
\hline
Date & Angle & Separation & Reference \\
(Bess. yr.) & (degrees) & (arcsec) & \\
\hline
1979.7420 & 125.4 & 0.300 & \cite{1980ATsir1097....3T}		\\
1984.7780 & 177.2 & 0.214 & \cite{1985AAS...61..483T}		\\
1991.2500 & 26.00 & 0.221 & \cite{1997ESASP1200.....E}\tablefootmark{a}		\\
1995.7593 & 65.70 & 0.343 & \cite{1997AJ....114.1639H}		\\
1996.4199 & 69.60 & 0.361 & \cite{2000AJ....119.3084H}		\\
1996.6903 & 72.10 & 0.355 & \cite{2000AJ....119.3084H}		\\
1997.5203 & 78.20 & 0.355 & \cite{1999AJ....117..548H}		\\
1997.5231 & 75.50 & 0.371 & \cite{1999AJ....117..548H}		\\
1999.8824 & 88.40 & 0.371 & \cite{2002AJ....123.3442H}		\\
2001.4991 & 98.00 & 0.354 & \cite{2011AJ....142..176M}		\\
2003.4840 & 108.6 & 0.340 & \cite{2004AJ....128.3012M}	\\
2003.5980 & 110.6 & 0.310 & \cite{2004AJ....128.3012M}	\\
2003.6500 & 110.2 & 0.340 & \cite{2004AJ....128.3012M}	\\
2003.7670 & 111.1 & 0.300 & \cite{2004AJ....128.3012M}	\\
2006.5615 & 132.8 & 0.286 & \cite{2009AJ....138..813H}		\\
\hline
\end{tabular}
\label{tab:speckle_data}
\tablefoot{%
\tablefoottext{a}{Hipparcos measure.}
}
\end{table}
%\end{table*}

\twocolumn

\clearpage

%%%%%%%%%%%%%%%%%%%%%%%%%%%%%%%%%%%%%%%%%%%%%%%%%%%%%%%%%%%%%
%%%%%%%%%%%%%%%%%%%%%%%%%%Observable model%%%%%%%%%%%%%%%%%%%%%%%%%%%
%%%%%%%%%%%%%%%%%%%%%%%%%%%%%%%%%%%%%%%%%%%%%%%%%%%%%%%%%%%%%

\section{Observable model}\label{app:obs-mod}

The radial velocities $V_A$ and $V_B$ and the coordinates of the orbit on the plane of the sky $(x,y)$ are calculated from the orbital parameters ${\cal P}_{\rm orb}=(P,T,V_0,K_A,K_B,e,\omega,a,i,\Omega)$ by means of the following equations:
\begin{equation}
V_A = V_0 + K_A\,[e\,\cos\omega + \cos(\nu+\omega)],
\end{equation}
\begin{equation}
V_B = V_0 - K_B\,[e\,\cos\omega + \cos(\nu+\omega)],
\end{equation}
\begin{equation}
x = AX + FY,
\end{equation}
\begin{equation}
y = BX + GY,
\end{equation}
where $V_0$ is the systemic velocity, $K_A$ and $K_B$ are the semi-amplitudes of the radial velocities for each component, $e$ is the orbital eccentricity and $\omega$ is the argument of periastron. The Thiele-Innes elements $A$, $B$, $F$ and $G$ are given by:
\begin{equation}
A = a\,(\cos\omega\cos\Omega - \sin\omega\sin\Omega\cos i),
\end{equation}
\begin{equation}
B = a\,(\cos\omega\sin\Omega + \sin\omega\cos\Omega\cos i),
\end{equation}
\begin{equation}
F = -a\,(\sin\omega\cos\Omega + \cos\omega\sin\Omega\cos i),
\end{equation}
\begin{equation}
G = -a\,(\sin\omega\sin\Omega - \cos\omega\cos\Omega\cos i),
\end{equation}
where $a$ is the semi-major axis of the relative orbit, $\Omega$ is the position angle of the line of nodes and $i$ is the inclination of the plane of the orbit to the plane of the sky. The true anomaly $\nu$ and the normalized rectangular coordinates in the true orbit ($X$, $Y$) are derived as follows:
\begin{equation}
\tan\frac{\nu}{2} = \sqrt{\frac{1+e}{1-e}} \, \tan\frac{E}{2},
\end{equation}
\begin{equation}
X = \cos E -e,
\end{equation}
\begin{equation}
Y = \sqrt{1-e^2}\, \sin E,
\end{equation}
where $E$ is the eccentric anomaly. This latter can be found using a fixed-point method to solve the Kepler's equation for any time~$t$:
\begin{equation}
\frac{2\pi}{P}\,(t-T) = E - e\,\sin E,
\end{equation}
where $P$ is the orbital period and $T$ is the time of periastron passage.

%%%%%%%%%%%%%%%%%%%%%%%%%%%%%%%%%%%%%%%%%%%%%%%%%%%%%%%%%%%%%
%%%%%%%%%%%%%%%%%%%%%%%%%%Bayesian approach%%%%%%%%%%%%%%%%%%%%%%%%%
%%%%%%%%%%%%%%%%%%%%%%%%%%%%%%%%%%%%%%%%%%%%%%%%%%%%%%%%%%%%%

\section{Bayesian approach}\label{app:bay-app}
For deriving the orbital parameters of HD~188753, we adopted a Bayesian approach as explained below.

According to Bayes’ theorem, the posterior probability of the orbital parameters ${\cal P}_{\rm orb}$ given the data D is stated as:
\begin{equation}
P({\cal P}_{\rm orb} | {\rm D}) = \frac{ P({\cal P}_{\rm orb}) \, P({\rm D} | {\cal P}_{\rm orb})}{P({\rm D})},
\end{equation}
where $P({\cal P}_{\rm orb})$ is the prior probability of the orbital parameters, $P({\rm D})$ is the global normalisation likelihood and $P({{\rm D} | \cal P}_{\rm orb})$ is the likelihood of the data given the orbital parameters, ${\cal L}$, defined in \refsec{sec:orb_meth}. The derivation of the posterior probabilities can be done using the Metropolis-Hastings algorithm (see, as a starting point, \citealt{2014aste.book..123A}). We used a Markov Chain for exploring the space to go from a set ${\cal P}_{\rm orb}^{t}$ to another set ${\cal P}_{\rm orb}^{t'}$, assuming that either set has the same probability, \ie $P({\cal P}_{\rm orb}^{t})=P({\cal P}_{\rm orb}^{t'})$. The Metropolis-Hastings algorithm requires that we compute the following ratio:
\begin{equation}
r=\frac{P({\cal P}_{\rm orb}^{t'} | {\rm D})}{P({\cal P}_{\rm orb}^{t} | {\rm D})}=\frac{P({\rm D} |  {\cal P}_{\rm orb}^{t'})}{P({\rm D} | {\cal P}_{\rm orb}^{t})},
\end{equation}
which is simply the ratio of the likelihood given in \refeq{eq:lik_global}. The acceptance probability of the new set, ${\cal P}_{\rm orb}^{t'}$, is then defined as:
\begin{equation}
\alpha({\cal P}_{\rm orb}^{t},{\cal P}_{\rm orb}^{t'}) = {\rm min}\,(1,r) = {\rm min}\left(1,\frac{P({\rm D} |  {\cal P}_{\rm orb}^{t'})}{P({\rm D} | {\cal P}_{\rm orb}^{t})}\right).
\end{equation}
The new values of the orbital parameters are accepted if $\beta \leq \alpha$, where $\beta$ is a random number drawn from a uniform distribution over the interval $[0,1]$, and rejected otherwise.

We set ten chains of 10 million points each with starting points taken randomly from appropriate distributions. The new set of orbital parameters is computed using a random walk as:
\begin{equation}
{\cal P}_{\rm orb}^{t'}={\cal P}_{\rm orb}^{t}+\alpha_{\rm rate} \, \Delta {\cal P}_{\rm orb},
\end{equation}
where $\Delta {\cal P}_{\rm orb}$ is given by a multinomial normal distribution with independent parameters and $\alpha_{\rm rate}$ is an adjustable parameter that is reduced by a factor of two until the rate of acceptance of the new set exceeds 25\%. We then derived the posterior probability of each parameter from the chains after rejecting the initial burn-in phase (\ie the first 10\% of each chain). For all parameters, we computed the median and the credible intervals at $16\%$ and $84\%$, corresponding to a 1-$\sigma$ interval for a normal distribution. The advantage of this percentile definition over the mode (maximum of the posterior distribution) or the mean (average of the distribution) is that it is conservative with respect to any change of variable over these parameters.

%%%%%%%%%%%%%%%%%%%%%%%%%%%%%%%%%%%%%%%%%%%%%%%%%%%%%%%%%
%%%%%%%%%%%%%%%%%%%%%%%%%%Radial-velocity solutions%%%%%%%%%%%%%%%%%%%%%%%%%%%
%%%%%%%%%%%%%%%%%%%%%%%%%%%%%%%%%%%%%%%%%%%%%%%%%%%%%%%%%

\begin{table*}[!h]
\section{Radial-velocity solutions}\label{app:orb-param}
\caption{Orbital solution for star~A using the radial-velocity measurements from \cite{1977Obs....97...15G}.}
\centering
\begin{tabular}{l c c c} 
\hline
\hline
Parameter & Median & 84$\%$ interval & 16$\%$ interval \\
\hline
$P$ (days) & 154.8 & +0.7 & -0.7 \\
$T$ (MJD-$40\,000$)\tablefootmark{a} & 2243 & +13 & -12 \\
$V_0$ (km s$^{-1}$) & -20.7 & +0.2 & -0.2 \\
$K_{A}$ (km s$^{-1}$) & 2.5 & +0.4 & -0.3 \\
$e$ & 0.21 & +0.12 & -0.12 \\
$\omega$ (degrees) & 56 & +35 & -30 \\
$a_1$ (km s$^{-1}$ yr$^{-1}$)\tablefootmark{b} & -0.34 & +0.13 & -0.13 \\
\hline
 & Star & $N_{\rm meas}$ & rms (m s$^{-1}$) \\
 & A\tablefootmark{c} & 54 & 841 \\
\hline
\end{tabular}
\tablefoot{%
\tablefoottext{a}{The Modified Julian Date (MJD) is defined as ${\rm MJD} = {\rm JD} - 2\,400\,000.5$.}
\tablefoottext{b}{The long-period orbital motion was taken into account through the linear term $a_1(t-T)$.}
\tablefoottext{c}{Given the strong line blending, the linear trend for star~A is affected by the 154-day modulation of the close pair.}
}
\end{table*}

\begin{table*}[!h]
\caption{Orbital solution for star~A using the radial-velocity measurements from \cite{2005Natur.436..230K}.}
\centering
\begin{tabular}{l c c c} 
\hline
\hline
Parameter & Median & 84$\%$ interval & 16$\%$ interval \\
\hline
$a_1$ (km s$^{-1}$ yr$^{-1}$) & -0.45 & +0.02 & -0.02 \\
\hline
 & Star & $N_{\rm meas}$ & rms (m s$^{-1}$) \\
 & A & 11 & 151 \\
\hline
\end{tabular}
\end{table*}

\begin{table*}[!h]
\caption{Orbital solution for star~Ba using the radial-velocity measurements from \cite{2005Natur.436..230K}.}
\centering
\begin{tabular}{l c c c} 
\hline
\hline
Parameter & Median & 84$\%$ interval & 16$\%$ interval \\
\hline
$P$ (days) & 153.97 & +0.07 & -0.07 \\
$T$ (JD-$2\,450\,000$) & 3245.9 & +0.7 & -0.8 \\
$V_0$ (km s$^{-1}$) & -22.24 & +0.06 & -0.06 \\
$K_{Ba}$ (km s$^{-1}$) & 13.30 & +0.04 & -0.04 \\
$e$ & 0.151 & +0.008 & -0.008 \\
$\omega$ (degrees) & 126.3 & +1.7 & -1.9 \\
\hline
 & Star & $N_{\rm meas}$ & rms (m s$^{-1}$) \\
 & Ba & 11 & 104 \\
\hline
\end{tabular}
\end{table*}

\begin{table*}[!h]
\caption{Orbital solution for star~A using the radial-velocity measurements from \cite{2009MNRAS.399..906M}.}
\centering
\begin{tabular}{l c c c} 
\hline
\hline
Parameter & Median & 84$\%$ interval & 16$\%$ interval \\
\hline
$a_1$ (km s$^{-1}$ yr$^{-1}$) & -0.60 & +0.06 & -0.06 \\
\hline
 & Star & $N_{\rm meas}$ & rms (m s$^{-1}$) \\
 & A & 35 & 98 \\
\hline
\end{tabular}
\end{table*}

\begin{table*}[!h]
\caption{Orbital solution for stars~Ba and~Bb using the radial-velocity measurements from \cite{2009MNRAS.399..906M}.}
\centering
\begin{tabular}{l c c c} 
\hline
\hline
Parameter & Median & 84$\%$ interval & 16$\%$ interval \\
\hline
$P$ (days) & 154.45 & +0.09 & -0.09 \\
$T$ (JD-$2\,450\,000$) & $3\,713.9$ & +0.4 & -0.4 \\
$V_0$ (km s$^{-1}$) & -21.61 & +0.03 & -0.03 \\
$K_{Ba}$ (km s$^{-1}$) & 13.48 & +0.03 & -0.03 \\
$K_{Bb}$ (km s$^{-1}$) & 17.56 & +0.13 & -0.13 \\
$e$ & 0.175 & +0.002 & -0.002 \\
$\omega$ (degrees) & 134.8 & +1.1 & -1.1 \\
$a_1$ (km s$^{-1}$ yr$^{-1}$) & 0.34 & +0.09 & -0.07 \\
\hline
$M_{Ba}\sin^3 i$ ($M_\odot$) & 0.258 & +0.004 & -0.004 \\
$M_{Bb}\sin^3 i$ ($M_\odot$) & 0.198 & +0.002 & -0.002 \\
\hline
 & Star & $N_{\rm meas}$ & rms (m s$^{-1}$) \\
 & Ba & 35 & 72 \\
 & Bb & 35 & 457\\
\hline
\end{tabular}
\label{tab:mazeh_B}
\end{table*}

%%%%%%%%%%%%%%%%%%%%%%%%%%%%%%%%%%%%%%%%%%%%%%%%%%%%%%%%%%%%%
%%%%%%%%%%%%%%%%%%%%%%%%%%Seismic data%%%%%%%%%%%%%%%%%%%%%%%%%%%
%%%%%%%%%%%%%%%%%%%%%%%%%%%%%%%%%%%%%%%%%%%%%%%%%%%%%%%%%%%%%

\begin{table*}[!h]
\section{Seismic data}\label{app:osc_freq}
\caption{Frequencies for star~A.}
\centering
\begin{tabular}{c c c} 
\hline
\hline
$l$ & Frequency & 1-$\sigma$ error \\
 & ($\mu$Hz) & ($\mu$Hz) \\
\hline
0 & 1770.42 & 0.18 \\
0 & 1875.97 & 0.13 \\
0 & 1982.17 & 0.07 \\
0 & 2088.81 & 0.12 \\
0 & 2196.04 & 0.09 \\
0 & 2303.19 & 0.09 \\
0 & 2410.03 & 0.10 \\
0 & 2517.29 & 0.23 \\
0 & 2625.52 & 0.48 \\
0 & 2732.74 & 0.33 \\        
\hline
1 & 1712.61 & 0.13 \\
1 & 1818.91 & 0.14 \\
1 & 1924.39 & 0.12 \\
1 & 2031.43 & 0.09 \\
1 & 2138.75 & 0.07 \\
1 & 2246.03 & 0.09 \\
1 & 2353.59 & 0.09 \\
1 & 2461.19 & 0.12 \\
1 & 2569.13 & 0.24 \\
1 & 2677.45 & 0.39 \\       
\hline
2 & 1976.76 & 0.12 \\
2 & 2083.76 & 0.11 \\
2 & 2191.54 & 0.10 \\
2 & 2298.92 & 0.14 \\
2 & 2405.95 & 0.12 \\
2 & 2513.76 & 0.31 \\
2 & 2622.08 & 0.82 \\
\hline
\end{tabular}
\tablefoot{The first column is the spherical harmonic degree. The second column is the temporal frequency. The third column is the 1-$\sigma$ uncertainty quoted when the mode is fitted. All modes were correctly detected and fitted. No correction for Doppler shifts was applied.}
\end{table*}

\begin{table*}[!h]
\caption{Frequencies for star~Ba.}
\centering
\begin{tabular}{c c c} 
\hline
\hline
$l$ & Frequency & 1-$\sigma$ error \\
 & ($\mu$Hz) & ($\mu$Hz) \\
\hline
0 & 3032.66 & 0.59 \\
0 & 3179.87 & 0.07 \\
0 & 3327.29 & 0.10 \\
0 & 3474.60 & 0.08 \\ 
\hline
1 & 2955.81 & 0.19 \\
1 & 3103.50 & 0.08 \\
1 & 3251.46 & 0.09 \\
1 & 3397.96 & 0.11 \\
1 & 3546.32 & 0.22 \\        
\hline
2 & 3172.14 & 0.08 \\
2 & 3320.03 & 0.08 \\
2 & 3466.91 & 0.47 \\      
\hline
\end{tabular}
\tablefoot{The first column is the spherical harmonic degree. The second column is the temporal frequency. The third column is the 1-$\sigma$ uncertainty quoted when the mode is fitted. All modes were correctly detected and fitted. No correction for Doppler shifts was applied.}
\end{table*}

\end{document}